\documentclass[11pt,a4paper]{article}
\pdfoutput=1
\usepackage[english]{babel}
\usepackage{amsfonts,amsbsy,bm,euscript,mathrsfs}
\usepackage{amssymb,stmaryrd,faktor,slashed}
\usepackage[normalem]{ulem}
\usepackage[x11names]{xcolor}
\usepackage[tbtags]{amsmath}
\usepackage[bookmarks=true,colorlinks=true,linkcolor=black,citecolor=black,urlcolor=black,bookmarksnumbered]{hyperref}
\usepackage[nosort]{cite}
\usepackage{tensor,braket,dsfont}

\newcommand{\abs}[1]{\ensuremath{\left\vert#1\right\vert}}
\newcommand{\dd}{\mathrm{d}}

\newcommand{\scom}[2]{\left[ #1 \stackrel{\star}{,} #2 \right]}
\newcommand{\tr}{\mathrm{tr}}

\numberwithin{equation}{section}

\newcommand{\hstar}{\, \hat{\star}\, }

\begin{document}

\vspace{36pt}

\begin{center}
{\huge{\bf Non-commutative deformations\\
of gauge theories\\
\vspace{8pt}
via Drinfel'd twists of the scale symmetry}} 

\vspace{36pt}

Riccardo Borsato and Tim Meier

\vspace{24pt}

{
\small {\it 
Instituto Galego de F\'isica de Altas Enerx\'ias (IGFAE),\\[2pt]
and Departamento de F\'\i sica de Part\'\i culas,\\[2pt]
Universidade de  Santiago de Compostela,\\[2pt]
15705 Santiago de Compostela,  Spain\\[4pt]}
\vspace{12pt}
\texttt{riccardo.borsato@usc.es}, \qquad \texttt{tim.meier@usc.es}}\\

\vspace{36pt}

{\bf Abstract}
\end{center}
\noindent
In this paper we consider  gauge theories that are relativistic and scale-invariant, and we construct their deformed versions via suitable star products. In particular, the non-commutative structure is controlled by Drinfel'd twists that are built out of symmetry generators that include the scale transformation. To achieve this, we construct a twisted differential calculus that allows us to identify the proper gauge-covariant quantities. We also show that our construction is equivalent to twists where the symmetry generators are implemented as active transformations of fields. As a  consequence of our construction,  the deformed gauge theories possess a twisted version of the original symmetry group. Moreover, at the planar level, the deformation is encoded just on the external legs of Feynman diagrams, leaving then the amputated diagrams undeformed. This work extends previous constructions and allows us to define twist-deformations of $\mathcal N=4$ super Yang-Mills that are conjectured to be holographically dual to a class of homogeneous Yang-Baxter deformations of $AdS_5\times S^5$.

\newpage 


\tableofcontents


\section{Introduction}

Non-commutative spacetimes and field theories have long been studied in the theoretical physics literature, their first motivation ranging back to the early stages of quantum field theory with the purpose of introducing an ultraviolet cutoff~\cite{Snyder:1946qz}. In the realm of quantum gravity, it was further argued that spacetime is likely to become non-commutative at the Planck scale, $[x^\mu,x^\nu]\sim \mathcal O(M_{Pl}^{-1})$~\cite{Doplicher:1994tu}, therefore providing an effective description of quantum gravity, see for example~\cite{Arzano:2021scz,Addazi:2021xuf}. A non-commutative gauge theory, in fact, emerges when considering the low-energy effective dynamics of open strings in the presence of a $B$-field~\cite{Chu:1999gi,Schomerus:1999ug}, and then these theories also appear  in the context of the AdS/CFT correspondence~\cite{Maldacena:1997re,Maldacena:1999mh,Hashimoto:1999ut}.
The original construction of non-commutative spacetime---that replaces the smooth manifold of spacetime in cartesian coordinates by operators with constant non-vanishing commutators~\cite{Snyder:1946qz}---can be mapped via the Weyl quantization to a smooth spacetime with ordinary fields, but with a non-commutative and associative product~\cite{Szabo:2001kg,Madore:2000en}. This product of fields is usually referred to as the star product and denoted by $\star$. While the original constant non-commutative spacetime corresponds to the Groenewold-Moyal star product, the ideas can be extended to a wide class of non-commutative spaces and star products \cite{Kontsevich:1997vb}. In the very general setup of star products, we will focus on star products generated by a Drinfel'd twist \cite{drinfeld_YBESolutions_1983}. Besides offering clear algebraic properties, a notion of twisted spacetime symmetries \cite{Wess:2003da,Chaichian:2004yh,Chaichian:2004za} and a differential calculus \cite{Aschieri:2009zz}, this class of non-commutative deformations of field theories is motivated also by the interest in homogeneous Yang-Baxter deformations in the context of the AdS/CFT correspondence   \cite{Vicedo:2015pna,vanTongeren:2015uha,Borsato:2021fuy,vanTongeren:2016eeb,Meier:2025tjq}, as we will mention below.

In principle, the twist induced by the star product can be generated by any set of vector fields $X_i$ acting on spacetime coordinates, e.g.~$f(x)\star g(x)=f(x)g(x)+X_1(f(x))X_2(g(x))+\ldots$, as long as certain desired properties, such as for example the associativity of $\star$, are ensured. Nevertheless, because of our motivations, here we will not consider generic vector fields $X_i$, but only those that implement \emph{symmetries} of the field theory that we wish to deform. At the same time, while in the context of non-commutative field theories these star products are constructed only out of generators of \emph{spacetime} transformations---therefore reformulating the non-commutative spacetime structure---here we will sometimes employ more general star products, that may involve also generators of \emph{internal} symmetries of the field theory.

In this paper, our motivation to consider non-commutative gauge theories, or more generally ``star-deformed'' gauge theories, is the fact that they naturally emerge from generalisations of the AdS/CFT correspondence~\cite{vanTongeren:2015uha,vanTongeren:2016eeb,Maldacena:1999mh,Hashimoto:1999ut} that are natural to consider given the integrability that appears in the planar limit of $\mathcal N=4$ super Yang-Mills~\cite{Beisert:2003yb,Minahan:2002ve}. On the one hand, when studying the low-energy limit of $D3$-branes in the presence of a constant $B$-field, for example, one recovers the Groenewold-Moyal deformation of $\mathcal N=4$ super Yang-Mills~\cite{Seiberg:1999vs,Douglas:2001ba,Szabo:2001kg}. This is arguably the simplest possible non-commutative deformation, since the non-commutativity is captured by a constant non-commutative parameter $\theta^{\mu\nu}$, so that $[x^\mu,x^\nu]=i\theta^{\mu\nu}$. Non-commutative gauge theories are therefore unavoidable if one turns on a background $B$-field, and for this reason it is natural to consider them also in the context of the AdS/CFT correspondence.
On the other hand, it is natural to ask whether the integrability of the planar $\mathcal N=4$ super Yang-Mills can be extended to deformations of the gauge theory. This question can be addressed more easily on the holographically-dual side, by constructing integrable deformations of the string sigma model on the $AdS_5\times S^5$ background. Here we are interested in a special class of integrable deformations that has been called homogeneous Yang-Baxter deformations~\cite{Vicedo:2015pna,vanTongeren:2015uha,Borsato:2021fuy}. Their classification is controlled by an ``$r$-matrix'' that solves the classical Yang-Baxter equation on the $\mathfrak{psu}(2,2|4)$ superalgebra. In~\cite{vanTongeren:2015uha,vanTongeren:2016eeb} it was argued that the duals of Yang-Baxter deformations of the string should be deformations of $\mathcal N=4$ super Yang-Mills obtained by replacing the ordinary product of fields with star products constructed from Drinfel'd twists which are, in fact, in one-to-one correspondence with $r$-matrices. Drinfel'd twists $\mathcal F$, which will be briefly reviewed in the main text, act on the tensor product of two copies of the space of functions, and they essentially allow to implement the star product by applying the standard product $f\cdot g=\mu(f,g)$ only after the implementation of the twist, i.e.~$f\star g=\mu(\mathcal F^{-1}(f,g))$. 

Although one can consider more complicated examples, like twists of Jordanian type, the simplest Drinfel'd twists are built out of a set of commuting generators. These ``abelian twists'' capture star products that include the above Groenewold-Moyal deformation, but also the light-like dipole deformation~\cite{Guica:2017mtd}, the angular dipole deformation\cite{Meier:2025tjq} and the $\beta$ and $\gamma$ deformations \cite{Lunin:2005jy}, which are explicit examples of the Leigh–Strassler deformation~\cite{Leigh:1995ep} which, although giving rise to an ordinary commutative gauge theory,  can be formulated in terms of a star product built out of $R$-symmetry generators. These abelian twists may be written as $e^{i \lambda r^{ij}Q_i\wedge Q_j}$, where $\lambda$ is a deformation parameter so that in the $\lambda\to 0$ limit the deformation is trivial, $r^{ij}$ are the components of the $r$-matrix, and $Q_i$ are symmetry generators satisfying $[Q_i,Q_j]=0$. In this paper we will not need to restrict ourselves to abelian twists, and the results will apply also to non-abelian ones.

The problem addressed in this paper is that for certain Yang-Baxter deformations an obstruction appears when trying to implement  the star products on the dual gauge theory side. This obstruction may be traced to the fact that, as mentioned above, the standard approach to constructing non-commutative field theories is to build the star product out of some given vector fields that act as Lie derivatives. For example, in the case of the Groenewold-Moyal deformation, one may consider the translations generated by $\partial_\mu$ and construct the star product as $f(x)\star g(x)=\lim_{y\to x}e^{i\theta^{\mu\nu}\frac{\partial}{\partial x^\mu} \frac{\partial}{\partial y^\mu}}f(x)g(y)=f(x)g(x)+i\theta^{\mu\nu}\partial_\mu f(x)\partial_\mu g(x)+\ldots$. The same logic can be applied to more general vector fields, and it was used, in fact, to construct non-commutative deformations of field theories even when the vector fields do not correspond to symmetries of the original field theory, see for example~\cite{Borowiec:2008uj}. However, issues may arise when attempting to construct non-commutative deformations of \emph{gauge} theories. According to the usual recipe, the gauge transformations for fields of the original theory should be again modified by replacing the ordinary products with star products, but it turns out that this can sometimes produce the breakdown of gauge invariance. This issue does not arise, for example, for the Groenewold-Moyal deformation, the dipole deformation of~\cite{Guica:2017mtd} and the Leigh-Strassler deformation, so that in these cases a deformation of the gauge theory can be straightforwardly constructed. But the issue is present whenever at least one symmetry generator $Q_i$ is implemented by a vector field $X_i$ that does not commute with partial derivatives, $[X_i,\partial_\mu]\neq 0$. This problem must be resolved, then, if one wants to construct all gauge theory duals of Yang-Baxter deformations.

Examples of vector fields that do not commute with partial derivatives are those implementing Lorentz transformations. For twists  including  Lorentz generators, the problem was addressed and solved in~\cite{Meier:2023lku} by providing a twisted differential calculus. The main idea to define gauge covariant objects was to reformulate the gauge (vector) fields in terms of differential forms. This allowed then to use the recipe of replacing products by star products, while ensuring gauge covariance \cite{Aschieri:2006ye}. This idea was used to construct gauge-invariant actions perturbatively in the deformation parameter for the $\kappa$-Minkowski case, see for example \cite{Dimitrijevic:2014dxa,Hersent:2022gry,Mathieu:2020ccc}. In the case of deformations of the Poincar\'e algebra, gauge invariant actions for finite deformation parameters have been constructed in~\cite{Meier:2023lku, Meier:2023kzt} including all kinds of matter fields. For the construction, it was necessary  to define a deformed notion of Hodge duality and to introduce the new concept of ``half forms'', that are useful to take into account also the Weyl and Dirac fermions in the theory. The reason for the success of that approach was essentially due to the fact that differential  forms (and the fermionic  half forms) transform just via Lie derivatives under Lorentz transformations.
Thanks to these results, it was possible to construct non-commutative gauge theories implementing all twists (even non-abelian) that may be constructed only out of Poincar\'e generators.

In this paper we want to go beyond~\cite{Meier:2023lku} by considering twists that include the scale transformation. Such twists naturally appear in the classification of all possible integrable deformations of AdS/CFT, since $\mathcal N=4$ is a conformal theory and the scale transformation is a symmetry that may be used to construct the twist. The simplest possible examples are abelian twists involving the scale transformation and generators that commute with it, for example internal symmetries or Lorentz generators, but non-abelian twists are also possible. We will start in section~\ref{sec:twist-diff-calc} with a generalisation of the construction of~\cite{Meier:2023lku} that encompasses also twists with the scale transformation. In order to do so, we will first need to introduce new objects that transform via Lie derivative also under the scale transformation, something which is not generally true for fields of gauge theories. Then, we will discuss the new notion of Hodge duality that we need, and all the technical tools that are necessary to implement the construction. In section~\ref{sec:gauge-theories} we will show how this formalism can be used   to easily construct the twist-deformed gauge theories. In section~\ref{sec:pedestrian} we will address the same problem in an alternative approach. First, we will show that if one insists on implementing the star product always with generators acting as Lie derivatives irrespectively of the nature of the field, then the definition of covariant derivative needs to be modified in order to respect gauge covariance. Second, we will also show that an equivalent interpretation of this construction is obtained by considering a star product where the symmetry generators act as ``active transformations'' on the fields, rather than Lie derivatives. We will also show that this interpretation is, in fact, equivalent also to the formulation in terms of the twisted differential calculus employed in~\cite{Meier:2023lku} and in the present paper. We will then discuss the twisted symmetries of the gauge theories in~\ref{sec:twist-symm}, and a planar equivalence theorem for Feynman diagrams in~\ref{sec:planar}. In the appendices we recap our conventions and present an alternative definition of a deformed Levi-Civita symbol.

In a related paper~\cite{next-paper} we will present the generalisation of the construction of twist-deformed gauge theories via star products, where the twists are built out of virtually any symmetry generators of the undeformed gauge theory.

\section{Twisted differential calculus}\label{sec:twist-diff-calc}
In the setup of twisted non-commutative theories, it is most natural to think of the twist acting via Lie derivatives on fields and functions, as well as on the volume form and differential forms. In order to construct a differential calculus that is compatible with gauge symmetry in the deformed setup, it is important to consider the correct objects when defining the gauge transformations. In fact,  for twists in the Poincar\'e algebra, it was shown in~\cite{Meier:2023lku} that the correct objects to work with are fields written as differential forms. First, being written as differential forms, the Lie derivatives with respect to vector fields act naturally on them. Second and more important for the following construction, given a Poincar\'e symmetric theory, the formulation via differential forms defines a good setup to compare between active transformations of the fields and passive transformations---where the underlying parametrization of the spacetime is transformed. We refer to appendix~\ref{app:conv} for our conventions. Before presenting the new construction, we give here a new perspective on the strategy of~\cite{Meier:2023lku} that will help understand the present paper.

If we consider for example a vector field $A_\mu$, its active transformation under a Poincar\'e transformation implemented by the vector field $\xi=\xi^\mu\partial_\mu$ (where $\xi^\mu$ may be spacetime-dependent) is given as
\begin{equation}
    A'_\mu(x)-A_\mu(x)\approx\delta_\xi A_\mu(x)=-\xi^\nu\partial_\nu A_\mu(x)-A_\nu(x)\partial_\mu\xi^\nu,
\end{equation}
where $\approx$ means that we only consider the infinitesimal transformation. Hence, on the differential form $A=A_\mu \dd x^\mu$ the active transformation acts as
\begin{equation}
    \delta_\xi A=\delta_\xi(A_\mu)\dd x^\mu=-\mathcal{L}_\xi(A).
\end{equation}
A complementary picture is the passive transformation generated by the coordinate change $x\rightarrow x'=x+\xi$, so that  on $A_\mu$ the passive transformation is
\begin{equation}
    A_\mu(x')-A_\mu(x)\approx\xi^\nu\partial_\nu A_\mu(x).
\end{equation}
While the former is naturally captured by a Lie derivative, naively it may seem that the latter is not. However, keeping in mind that a  passive transformation also affects the basis of differential forms as
\begin{equation}
\dd x^\mu\rightarrow\dd x'{}^{\mu}=\dd x^\mu+\dd\xi^\mu=(\delta_\nu^\mu+\partial_\nu\xi^\mu)\dd x^\nu,
\end{equation}
the combined one-form $A=A_\mu \dd x^\mu$  transforms again with the correct Lie derivative:
\begin{equation}
    A(x')-A(x)=\mathcal{L}_\xi(A(x)).
\end{equation}
Notice the relative minus sign between active and passive transformations, see also appendix~\ref{app:conv}.

When deforming by star products, the product rule for partial derivatives, as vector fields, is deformed, but the exterior derivative still acts via the usual Leibniz rule. Hence, in the setup of non-commutative gauge theories, in this formalism it is the exterior derivative that will be naturally covariantised by a gauge one-form. Accordingly, it is more natural to define the gauge transformation rule for the gauge one-form $A$ directly, rather than for its individual component fields $A_\mu$, cf. section~\ref{sec:gauge-theories}. 
This example gives us a guiding principle to construct objects like the one-form $A$ that transform in the same way (up to a sign) under passive and active transformations, and to assign them the gauge transformation rules in the deformed theory.
 In the case of twists in the Poincar\'e algebra, this for example led to the introduction of Grassmann-odd basis spinors in order to define gauge-invariant actions including fermionic matter fields \cite{Meier:2023lku}. 
Here we will further extend this framework to include deformations beyond those generated by the Poincaré algebra, by incorporating the scaling transformation. Since our approach relies on actively transforming the fields, the construction will typically yield consistent gauge-invariant deformed actions only for theories that are, at least classically, scale invariant. We refer to section \ref{sec:gauge-theories} for further comments on this.

While for Poincar\'e transformations it is enough to repackage the fields into differential forms in order to have an object that under (active) transformations  transforms purely by a Lie derivative, this is not enough for the scale transformation. For example, a scalar field $\phi$ transforms under scale transformations $x'{}^\mu=e^\epsilon x^\mu$ as 
\begin{equation}
\phi'(x)-\phi(x)\approx i\epsilon\ \delta_D(\phi(x))= i\epsilon(-\mathcal{L}_{D}(\phi(x))+i\Delta_\phi\phi(x)),
\end{equation}
where $D=-ix^\mu\partial_\mu$ and $\Delta_\phi =1$ is its scaling dimension in four spacetime dimensions. As in the example of Poincar\'e transformations, the passive transformations of $\phi$ are simply captured by a transformation of the spacetime dependence. Hence
\begin{equation}
    \phi(x')-\phi(x)\approx i\epsilon\, D(\phi)=\epsilon\,  x^\mu\partial_\mu\phi(x).
\end{equation}
In order to mimic the role that was played by  basis forms $\dd x^\mu$ in the case of Poincar\'e transformations, we introduce an auxiliary object $H$ that will serve as a ``basis scale'' with respect to scale transformations. It might be possible to interpret $H$ as a ``constant auxiliary field'', or as a new ``coordinate'' keeping track of the scale in the space of fields, but in our construction $H$ will be just a bookkeeping device and we will not need to assign to it a particular meaning.  We can now define a new field $\varphi=\phi H^{-1}$ and, taking into account that we want active transformations to leave $H$ invariant as it was the case for $\dd x^\mu$, we declare that $\varphi$ transforms as
\begin{equation}
    \delta_D\varphi=\delta_D(\phi)H^{-1}=-\mathcal{L}_D\varphi+i\varphi.
\end{equation}
Similar to the case of Poincar\'e transformations, we can now change the perspective and consider the field $\varphi$ as the fundamental field. If we want active and passive scale transformations to act on $\varphi$ in the same way up to a sign, we must extend the action of the passive transformation  so that it acts also on $H$. In the rest of the paper, we will use a new notation $\mathcal D$ for the ``improved'' passive transformation acting also on $H$, so that it satisfies
\begin{equation}
    \mathcal{D}H=i H,\qquad
    \mathcal{D}\phi=\mathcal{L}_D\phi\quad\implies\quad
    \mathcal{D}(\varphi)=\mathcal{L}_D\varphi-i\varphi.
\end{equation}
This also means that we can formally identify the operator $\mathcal D$ with
\begin{equation}
    \mathcal D = \mathcal L_D+i\, H\partial_H,
\end{equation}
where $\mathcal L_D$ is simply the Lie derivative with respect to $-ix^\mu\partial_\mu$.
Having found a rescaled field, $\varphi$, that transforms in the same way (up to a sign) under active and passive scale transformations, we then decide to work with $\varphi$ and, whenever we have to construct a twist involving a scale transformation, we use $\mathcal D$. This follows the natural perspective expected from the general philosophy of non-commutative gauge theories, where one would like to twist the structure of the spacetime and therefore wants to have a non-trivial action on coordinates. 

\vspace{12pt}

In the remainder of this subsection, we will show how scale invariant actions can be reformulated consistently after introducing the auxiliary field $H$. This  will be a preparation to consistently deform such actions with twists depending on $\mathcal D$.
Let us consider the action of a free \emph{massless} scalar field as the simplest example on how to incorporate the auxiliary field $H$. Starting from the standard kinetic action, we can insert the right powers of $H$ as
\begin{equation}
\label{eq:scalarCompUndeformed}
S=\frac{1}{2}\int \mathrm{d}^4x~\partial_\mu\phi\partial^\mu\phi=\frac{1}{2}\int \mathrm{d}^4x\ H^2~\partial_\mu\varphi\partial^\mu\varphi.
\end{equation}
This already prepares the action in the desired form with respect to scale transformations, but at the same time we also want to rephrase it in terms of differential forms in order to be compatible with deformations of the Poincar\'e algebra as well. A crucial object for a differential calculus in terms of differential forms is the Hodge operator. In order to construct a twist-deformed gauge-invariant action, it is crucial to have a notion of Hodge operator that commutes with the algebra generators employed in the twist \cite{Meier:2023lku,Meier:2023kzt}. The dilatation operator, however, does not commute with the Hodge operator, since the dilatation operator measures the degree of the form, and in general this degree may differ between a form and its Hodge dual. In particular, one has $[\mathcal L_D,*]\omega=-i(n-2k)*\omega$. However, we can use the auxiliary field $H$ to define a new duality between forms, one that does not change the scaling dimensions. In particular, the operator defined by
\begin{equation}
\label{eq:undefHodge}
*\left( \mathrm{d}x^{\mu_1}\wedge\dots\wedge \mathrm{d}x^{\mu_k} \right)=\frac{1}{(n-k)!}H^{n-2k}\tensor{\varepsilon}{^{\mu_1\dots\mu_k}_{\mu_{k+1}\dots\mu_{n}}}\dd x^{\mu_k+1}\wedge\dots\wedge\dd x^{\mu_n},
\end{equation}
shares similar properties with the usual Hodge operator. In fact, the new Hodge operator defines a duality between $k$-forms and $(n-k)$-forms in $n$ dimensions as it satisfies the properties
\begin{equation}
\begin{aligned}
**\omega&=-(-1)^{k(n-k)}\omega\\
*(f\omega)&=f*\omega\\
\int\omega\wedge*\chi&=\int \chi\wedge*\omega,
\end{aligned}
\end{equation}
where $\omega$ and $\chi$ are $k$-forms and $f$ is a function, or in other words a 0-form. The key difference between this Hodge operator and the usual one lies in the fact that it commutes with scaling transformations, i.e.
    \begin{equation}
    \begin{aligned}
        \mathcal{D}(*\omega)&=\mathcal{L}_D(*\omega)+iH\partial_H(*\omega)\\
        &=*(\mathcal{L}_D\omega)-i(n-2k)*\omega+i*H\partial_H\omega+i(n-2k)*\omega\\
       & =*(\mathcal D\omega).
    \end{aligned}
\end{equation}

We will continue referring to the new operator as the Hodge operator, and when we want to make sure that it is not confused with the usual one we will call it ``generalised'' Hodge operator. In the remainder of the paper, instead of the standard one, we will  use the new Hodge operator. We will see that in the context of the twist-deformed gauge theories, this generalised Hodge operator will also need to be deformed.

For the construction of the deformations later on, it will actually be convenient to define scale invariant basis one-forms $\mathrm{d}\xi^\mu$ as\footnote{In the deformed setup, this definition will need to be modified by the use of the star product.}
\begin{equation}\label{eq:undefScaleInvOneForm}
    \mathrm{d}\xi^\mu=H\mathrm{d}x^\mu.
\end{equation}
By multiplying our definition of the Hodge dual on basis forms in \eqref{eq:undefHodge}, we can directly derive a Hodge operator acting on scale invariant basis forms as\footnote{The extra signs compared to the previous formula are due to a reorganisation of indices in the $\varepsilon$-tensor. We prefer to write the formula in this way, now, because it will be easier to generalise it to the twisted setup.}
\begin{equation}
\label{eq:scaleInvHodgeUndeformed}
    *\left( \mathrm{d}\xi^{\mu_1} \wedge\dots \wedge \mathrm{d}\xi^{\mu_k} \right)=\frac{(-1)^{\sigma(k)}}{(n-k)!}\tensor{\varepsilon}{_{\mu_n\dots\mu_{k+1}}^{\mu_1\dots\mu_k}}\mathrm{d}\xi^{\mu_{k+1}}\wedge\dots\wedge\mathrm{d}\xi^{\mu_n},
\end{equation}
where $\sigma(k)=\frac{(n-k)(n+k-1)}{2}$. This formally takes the same definition as the usual Hodge operator, with the replacement $\dd x^\mu\to\dd \xi^\mu$. From now on we will always work with differential forms in the basis $\dd \xi^\mu$. In terms of the new Hodge operator, the scalar field action in \eqref{eq:scalarCompUndeformed} becomes
\begin{equation}
S=\frac{1}{2}\int \dd \varphi\wedge*\dd \varphi.
\end{equation}
This action is then identical to the one for $\phi$ that we started with, upon the replacement $\phi\to \varphi$, as well as the implicit one $\dd x^\mu\to\dd \xi^\mu$. However, the important point is that now all fundamental objects appearing in the action transform equivalently under active and passive transformations, both for Poincar\'e and for the scale transformation. This is then the perfect starting point for the deformation procedure.

\vspace{12pt}

In order to define a field theory action on non-commutative spaces, a differential calculus adapted to the twisted algebra is needed. As vector fields, also partial derivatives act via twisted coproducts on products of fields. In general, we can let the twist act via Lie derivatives on differential forms, which naturally introduces a twisted structure there as well. The resulting differential calculus has been previously discussed in \cite{Aschieri:2006ye,Aschieri:2009ky,Aschieri:2005zs}. In \cite{Meier:2023kzt,Meier:2023lku}, a deformation of the Hodge duality has been proposed in order to give a compatible structure for gauge invariant actions. In particular, for gauge invariance it was crucial that the Hodge operator is star-linear.

In this section, we will first review the construction of a differential calculus by introducing a twisted exterior algebra of forms, an exterior derivative, and a notion of integration. We will further review a notion of complex conjugation on star-forms and will introduce a metric tensor  to raise and lower indices. All these notions and properties were already used in~\cite{Meier:2023lku} for twists in the Poincar\'e algebra, and they automatically extend to our case that includes scale transformations thanks to the introduction of $H$ and of the new Hodge operator. Then, we will show that the metric tensor is not star commutative once we include the dilatation generator in the twist, which is then a new feature compared to deformations only in the Poincar\'e algebra. Furthermore, we will introduce a deformation of the Hodge operator that was defined above, and we will show that it is star-linear. In the last part of this section, we will define a deformed coderivative and a deformed Laplace operator, which we will show to be equal to the undeformed one. 

\subsection{Preliminary review on deformed differential calculus}
\label{sec:DefDiffCalc}

First, we closely follow the presentation of~\cite{Meier:2023lku} showing that the construction of that paper can be straightforwardly generalised to our case. We define the star product via a Drinfel'd twist that we denote by $\mathcal F$. This is a map $\mathcal F:V\otimes V\to V\otimes V$, where $V$ is a vector space carrying a representation of a Lie algebra. In our case, the Lie algebra under consideration is generated by the Poincar\'e transformations and the scale transformations. We refer to appendix~\ref{app:conv} for our conventions. If the ordinary product of two functions is implemented by $\mu:(g_1,g_2)\to g_1g_2$, star products of functions are then defined as
\begin{equation}
    g_1(x)\star g_2(x) = \mu(\bar{\mathcal F}(g_1(x),g_2(x))),
\end{equation}
where $\bar{\mathcal F}=\mathcal F^{-1}$ is the inverse twist. Sometimes we will also use the Sweedler notation $\bar{\mathcal F}(g_1,g_2)=\bar f^\alpha(g_1)\otimes \bar f_\alpha(g_2)$. The star product is non-commutative but associative, as a consequence of the properties defining Drinfel'd twists. See for example section 2 of~\cite{Meier:2023lku} for a small introduction to Drinfel'd twists following our conventions.

\paragraph{Twisted wedge product} We generalise the star product of functions (0-forms) to a twisted exterior product, the twisted wedge product, which is defined by
\begin{equation}
\omega\wedge_\star\chi=\bigwedge(\bar{\mathcal{F}}(\omega\otimes\chi))=\bar{f}^\alpha(\omega)\wedge\bar{f}_\alpha(\chi),
\end{equation}
where $\bar {\mathcal{F}}$ is the inverse twist. 

By applying the definition of the twist, one can always relate the twisted wedge products to the untwisted ones. In fact, remembering that the twist acts as Lie derivatives, when acting on basis one-forms we simply evaluate the corresponding leg of the twist in the appropriate representation. Following the conventions in \cite{Meier:2023lku}, this defines
\begin{equation}
\label{eq:defineF2}
\begin{aligned}
\dd x^\mu\star f&=\tensor{\bar{F}}{_\nu^\mu}f\dd x^\nu\\
f\star\dd x^\mu&=\tensor{\left(\bar{F}_{op}\right)}{_\nu^\mu}f\dd x^\nu\\
\dd x^\mu\wedge_\star\dd x^\nu&=\tensor{\bar{F}}{_\rho^\mu_\sigma^\nu}\dd x^\rho\wedge\dd x^\sigma,
\end{aligned}
\end{equation}
where $\tensor{\bar{F}}{_\nu^\mu},\tensor{\left(\bar{F}_{op}\right)}{_\nu^\mu}$ still acts as differential operators on the function $f$. We refer to \cite{Meier:2023lku} for more details and examples.\\

For a given differential form, we can express the coefficients with respect to a basis with or without star products, e.g.~for a one form:
\begin{equation}
\omega=\omega_\mu\dd x^\mu=\omega^\star_\mu\star\dd x^\mu,
\end{equation}
where according to \eqref{eq:defineF2} the two sets of coefficients must be related as
\begin{equation}
\omega^\star_\mu=\tensor{\left(F_{op}\right)}{_\mu^\nu}\omega_\nu.
\end{equation}

\paragraph{$\mathcal R$-matrix} In the setting of a (twisted) Hopf algebra, we can relate  the coproduct $\Delta$ and the opposite coproduct $\Delta_{op}=\tau \circ\Delta$, where $\tau$ is the permutation operator, by an $\mathcal R$ matrix as
\begin{equation}
    \Delta_{op}(X)=\mathcal{R}\Delta(X)\bar{\mathcal{R}},
\end{equation}
where again, $\bar{\mathcal{R}}$ denotes the inverse of the $\mathcal R$-matrix. Consequently, the $\mathcal R$-matrix relates a star product between a set of objects to the star product with the order reversed. Following again the conventions in \cite{Meier:2023lku}, moving a function over a basis form leads to
\begin{equation}
\begin{aligned}
\dd x^\mu\star f&=\bigwedge\left( \bar{\mathcal{F}}_{op} \mathcal{R}(\dd x^\mu \otimes f) \right)\\
&=\bigwedge\left( \bar{\mathcal{F}}\bar{\mathcal{R}}(f \otimes \dd x^\mu) \right)\\
&=\left(\bar R^\alpha f\right)\star\left(\tensor{(\bar R_\alpha)}{_\nu^\mu}\right)\mathrm{d}x^\nu\\
&=\left( \tensor{R}{_\nu^\mu}f \right)\star \dd x^\nu,
\end{aligned}
\end{equation}
for a given function $f$.\footnote{\label{foot:constantR}For completeness, let us stress that the representation of the $\mathcal R$-matrix with multiple indices cannot be defined for all possible twists. In particular, $\tensor{(\bar R_\alpha)}{_\nu^\mu}$ has to commute with the twist itself. However, as discussed in \cite{Meier:2023lku}, the latter is at least satisfied for twists constructed solely from vector fields that are linear in the coordinates, which is satisfied for all elements of the Poincar\'e algebra and the dilatation generator.}  Similarly, the star-wedge product between basis forms is $R$-antisymmetric as
\begin{equation}
\label{eq:R-antisym}
\dd x^\mu\wedge_\star \dd x^\nu=-\tensor{R}{_\rho^\mu_\sigma^\nu}\dd x^\sigma\wedge_\star \dd x^\rho=-\tensor{\bar{R}}{_\sigma^\nu_\rho^\mu}\dd x^\sigma\wedge_\star \dd x^\rho,
\end{equation}
where $\tensor{\bar{R}}{_\sigma^\nu_\rho^\mu}$ is the representation of the $\mathcal R$-matrix that acts with its two legs on different basis one-forms. Note that $\tensor{R}{_\rho^\mu}$ and $\tensor{R}{_\rho^\mu_\sigma^\nu}$ are related by evaluating the former on yet another basis one-form as
\begin{equation}
    \tensor{R}{_\rho^\mu}(\mathrm{d}x^\nu)=\tensor{R}{_\rho^\mu_\sigma^\nu}\mathrm{d}x^\sigma.
\end{equation}

\paragraph{Exterior derivative} Beyond the exterior algebra of differential forms above, we need to define an exterior derivative for a twisted differential calculus. As the twist is taken to act via Lie derivatives, which commute with the standard exterior derivative by definition, one concludes that the exterior derivative follows the usual Leibniz rule
\begin{equation}
\dd\left( \omega \wedge_\star \chi \right)=\dd\omega \wedge_\star \chi + (-1)^{\abs{\omega}}\omega \wedge_\star \dd\chi.
\end{equation}
It will be useful to express the exterior derivative in a basis via star products:
\begin{equation}
\dd f=\partial_\mu f\dd x^\mu=\partial^\star_\mu f \star \dd x^\mu,
\end{equation}
where $\partial_\mu^\star=\tensor{\left(\bar{F}_{op}\right)}{_\mu^\nu}\partial_\nu$. Furthermore, closedness of the exterior derivative leads to a commutation relation of partial star-derivatives \cite{Meier:2023lku}
\begin{equation}
\partial_\mu^\star\partial_\nu^\star=\tensor{R}{_\mu^\rho_\nu^\sigma}\partial_\sigma^\star\partial_\rho^\star=\tensor{\bar{R}}{_\nu^\sigma_\mu^\rho}\partial_\sigma^\star\partial_\rho^\star.
\end{equation}

\paragraph{Integration} The non-commutative product of course means that, in general, the star wedge product is not antisymmetric any more. For the construction of gauge invariant theories that we will perform later, it is desirable to have a cyclic star product under integration, i.e.
\begin{equation}
\label{eq:intCyc}
\begin{aligned}
\int \omega \wedge_\star \chi = (-1)^{\abs{\omega}\abs{\chi}}\int \chi \wedge_\star \omega + \text{total derivative},
\end{aligned}
\end{equation}
where $\abs{\omega}+\abs{\chi}=n$. Starting from the left-hand-side, we can use the Sweedler notation for the inverse twist $\bar{\mathcal{F}}=\bar f^\alpha \otimes\bar f_\alpha$ as \cite{Aschieri:2009ky}
\begin{equation}
\begin{aligned}
    \int \bar f^\alpha \omega\wedge\bar f_\alpha\chi=\int \bar f^\alpha_{(1)}(\omega\wedge S(\bar f^\alpha_{(2)})\bar f_\alpha\chi)=\int\omega\wedge S(\bar f^\alpha)\bar f_\alpha\chi+\int \bar{f}^{'\alpha}_{(1)}(\omega\wedge S(\bar f^{'\alpha}_{(2)})\bar f_\alpha\chi),
\end{aligned}
\end{equation}
where $S$ is the antipode of the undeformed Hopf algebra, which maps elements of the Lie algebra to its negative, i.e. $S(X)=-X$, and is extended to the full Hopf algebra antimultiplicatively, $S(XY)=S(Y)S(X)$. Furthermore, we used the Sweedler notation for the undeformed coproduct, i.e. $\Delta(X)=X_{(1)}\otimes X_{(2)}$, and in the last equation the coproduct was expressed explicitly, such that $\bar{f}^{'\alpha}_{(1)}\neq 1$. In the case of $\bar{f}^{'\alpha}_{(1)}$ being part of the Poincar\'e algebra, the last term becomes an integral over a total Lie derivative of a top form, which is a total derivative. In the case of $\bar{f}^{'\alpha}_{(1)}$ containing the generator $\mathcal{D}$, this is only the case if $\int \omega\wedge S(\bar f^{'\alpha}_{(2)})\bar f_\alpha\chi$ is overall independent of $H$, i.e. it is scale invariant. Hence, 
\begin{equation}
    \int \omega\wedge_\star\chi=\int\omega\wedge S(\bar f^\alpha)\bar f_\alpha\chi+\text{total derivative},
\end{equation}
for all scale invariant terms. Similarly, the right-hand-side of \eqref{eq:intCyc} becomes
\begin{equation}
    (-1)^{\abs{\omega}\abs{\chi}}\int \chi \wedge_\star \omega=\int S(\bar f^\alpha)\bar f_\alpha\omega\wedge\chi+\text{total derivative}.
\end{equation}
Thus, equation \eqref{eq:intCyc} holds if the twist satisfies what we will call the ``twist unimodularity condition'' \cite{Aschieri:2009ky}
\begin{equation}\label{eq:TwistUnimodularity}
\begin{aligned}
S\left( f^\alpha \right)f_\alpha=1,
\end{aligned}
\end{equation}
or equivalently
\begin{equation}
S\left( f_\alpha \right)f^\alpha=1,\qquad
S\left( \bar{f}^\alpha \right)\bar{f}_\alpha=1,\qquad
S\left( \bar{f}_\alpha \right)\bar{f}^\alpha=1.
\end{equation}
This restriction can be translated to a condition on the $\mathcal R$-matrix as
\begin{equation}
\begin{aligned}
\bar{R}^\alpha \bar{R}_\alpha=S(R^\alpha)R_\alpha=S(\bar{f}^\beta)S(f_\alpha)f^\alpha f_\beta=1.
\end{aligned}
\end{equation}
The latter equation acting on a basis one form results in
\begin{equation}
\begin{aligned}
\dd x^\mu&=\bar{R}^\alpha\bar{R}_\alpha(\dd x^\mu)\\
&= \tensor{R}{_\nu^\mu}(\dd x^\nu)\\
&= \tensor{R}{_\nu^\mu_\rho^\nu}\dd x^\rho
\end{aligned}
\end{equation}
and hence
\begin{equation}
\label{eq:RMatContraction}
\begin{aligned}
\tensor{R}{_\nu^\mu_\rho^\nu}=\delta_\rho^\mu.
\end{aligned}
\end{equation}

\paragraph{Conjugation} To construct field theory actions, we will need to have a way to promote complex conjugation from functions and one-forms to higher forms obtained via star-wedge products. In particular, we will require that the conjugate of a star-wedge product is
\begin{equation}
\overline{\omega \wedge_\star \chi} = (-1)^{\abs{\omega}\abs{\chi}}\overline{\chi}\wedge_\star\overline{\omega}.
\end{equation}
In general, the above puts constraints on the twist in the form of reality conditions. We will always work with real deformation parameters.


\subsection{Metric tensor}
Similarly to \cite{Meier:2023lku}, we can define a deformed metric tensor. At this point, new features appear that are special for twists involving the scale transformation.
We define a deformed metric tensor by using the usual undeformed constant metric $\eta_{\mu\nu}$ as
\begin{equation}
\eta^\star=\eta_{\mu\nu}\dd x^\mu\otimes_\star \dd x^\nu.
\end{equation}
By construction, this is invariant under Poincar\'e transformations and  it has a scaling dimension of $-2$. Hence, we can move a function through the metric tensor at the cost of acting with a new operator\footnote{Similar to the $\mathcal R$-matrix representations with several indices already used, $\tilde{R}$ denotes the representation of the $\mathcal R$-matrix that has one leg being evaluated on a Poincar\'e-invariant basis object of scaling dimension $-1$. As the metric is Poincar\'e invariant with scaling dimension $-2$, we have $\tilde R^{-2}$ appearing. Note that $\tilde R$ still acts as an operator, because the  second leg of the $\mathcal R$-matrix remains unevaluated. When acting with $\tilde R$ on a basis one-form, it will be represented as a matrix with two vector indices $\tensor{\tilde R}{_\mu^\nu}$, not to be confused with the formerly introduced operator $\tensor{R}{_\mu^\nu}$, which instead represents the $\mathcal R$-matrix acting with one leg on a basis one-form and its second leg being unevaluated.}  that we call $\tilde{R}$
\begin{equation}
\begin{aligned}
f\star \eta^\star=\eta^\star \star \tilde{R}^{-2}f.
\end{aligned}
\end{equation}
Equivalently, the function can be moved through each one-form individually as
\begin{equation}
\begin{aligned}
f\star \eta^\star&=\eta_{\mu\nu} f \star \dd x^\mu \otimes_\star \dd x^\nu\\
&=\eta_{\mu\nu} \dd x^\rho \otimes_\star \dd x^\sigma \star \tensor{\bar{R}}{_\sigma^\nu}\tensor{\bar{R}}{_\rho^\mu}f,
\end{aligned}
\end{equation}
and hence,
\begin{equation}
\begin{aligned}
\eta_{\mu\nu}\tilde{R}^{-2}&=\eta_{\rho\sigma}\tensor{\bar{R}}{_\nu^\rho}\tensor{\bar{R}}{_\mu^\sigma},
\end{aligned}
\end{equation}
or equivalently,
\begin{equation}
\label{eq:lowerRaiseRMat}
\tensor{R}{_\mu^\nu}\tilde{R}^{-2}=\tensor{\bar{R}}{^\nu_\mu},\qquad\qquad
\tilde{R}^{-2}\tensor{R}{_\mu^\nu}=\tensor{\bar{R}}{^\nu_\mu}.
\end{equation}
By star-multiplying from the right and applying the commutation relation of $H$ with basis forms, this leads to
\begin{equation}
\label{eq:MetricSimp1}
\begin{aligned}
\tensor{\left(\tilde{R}^{-1}\right)}{_\rho^\nu}\eta_{\mu\nu} \dd x^\mu \otimes_\star \dd x^\rho \star H^2=\eta_{\mu\nu} (\dd x^\mu \star H)\otimes_\star (\dd x^\nu \star H).
\end{aligned}
\end{equation}
As $\dd x^\mu \star H$ is scaleless and $(\dd x^\mu \star H) \otimes (\dd x_\mu \star H)$ is Poincar\'e invariant, it follows that
\begin{equation}
\begin{aligned}
\mathcal{D}(\dd x^\mu \star H) \otimes (\dd x_\mu \star H) &=(\dd x^\mu \star H) \otimes S(\mathcal{D})(\dd x_\mu \star H) \\
\mathcal{L}_X(\dd x^\mu \star H) \otimes (\dd x_\mu \star H) &=(\dd x^\mu \star H) \otimes \mathcal{L}_{S(X)}(\dd x_\mu \star H),
\end{aligned}
\end{equation}
where $X$ denotes any element of the Poincar\'e algebra and $S$ is the antipode of the Hopf algebra that is obtained by complementing the Poincar\'e symmetries with the scale transformation. Notice that in the above formulas the standard tensor product has been used. However, thanks to these observations,  equation \eqref{eq:MetricSimp1}  can be simplified as
\begin{equation}
\begin{aligned}
\eta_{\mu\nu} (\dd x^\mu \star H)\otimes_\star (\dd x^\nu \star H) &= \bar{f}^\alpha(\dd x^\mu \star H) \otimes \bar{f}_\alpha(\dd x_\mu \star H)\\
&= f^\alpha S(\bar{f}_\alpha)(\dd x^\mu \star H)\otimes (\dd x_\mu \star H)\\
&= (\dd x^\mu \star H)\otimes (\dd x_\mu \star H)\\
&= \eta_{\mu\nu} \tensor{\bar{\tilde{F}}}{_\rho^\mu}\tensor{\bar{\tilde{F}}}{_\sigma^\nu} (\dd x^\rho H)\otimes (\dd x^\sigma H),
\end{aligned}
\end{equation}
where $\tensor{\bar{\tilde{F}}}{_\sigma^\nu}$ is the representation of $\bar{\mathcal{F}}$ acting with its second leg on a Poincar\'e invariant object of scaling dimension $-1$, such as $H$, and with its first leg on a basis one-form. Furthermore, taking into account that $\tensor{\bar{\tilde{F}}}{_\rho^\mu}$  is the matrix of a Poincar\'e transformation, it follows that $\eta_{\mu\nu} \tensor{\bar{\tilde{F}}}{_\rho^\mu}\tensor{\bar{\tilde{F}}}{_\sigma^\nu}=\eta_{\rho\sigma}$. Hence, 
\begin{equation}
    \eta_{\mu\nu}(\dd x^\mu\star H)\otimes_\star(\dd x^\nu\star H)=\eta_{\mu\nu}(\dd x^\mu H)\otimes(\dd x^\nu H)=\eta H^2.
\end{equation}
Moreover, the undeformed metric tensor $\eta=\eta_{\mu\nu}\dd x^\mu\otimes\dd x^\nu$ is also Poincar\'e invariant and hence
\begin{equation}
\begin{aligned}
\tensor{\left(\tilde{R}^{-1}\right)}{_\rho^\nu}\eta_{\mu\nu} \dd x^\mu \otimes_\star \dd x^\rho \star H^2=\eta H^2=\eta \star H^2 = (\eta_{\mu\nu} \tensor{F}{_\rho^\mu_\sigma^\nu} \dd x^\rho \otimes_\star \dd x^\sigma)\star H^2.
\end{aligned}
\end{equation}
To conclude, by comparing the coefficients with respect to the star tensor products we obtain
\begin{equation}
\begin{aligned}
\eta_{\rho\sigma} \tensor{F}{_\mu^\rho_\nu^\sigma}=\tensor{\left(\tilde{R}^{-1}\right)}{_\nu^\rho}\eta_{\mu\rho},
\end{aligned}
\end{equation}
which will be useful later on in the discussion on the deformed Laplacian operator.

The latter equation is quite remarkable as it relates the twist evaluated on basis forms to the $\mathcal R$-matrix evaluated on $H$ and a basis form, both appropriately contracted with the metric. Note that in the case of a twist only of the Poincar\'e algebra, where $\tensor{\left(\tilde{R}^{-1}\right)}{_\nu^\rho}=\delta^\rho_\nu$, this identity reduces to the condition for the metric remaining undeformed. In section \ref{sec:Laplacian}, the latter identity will ensure undeformedness of the Laplace operator, which is a crucial property for free field theories in the undeformed and deformed setups to coincide.

\subsection{Hodge duality}
For a consistent gauge-invariant formulation of twisted non-commutative gauge theory, the notion of Hodge duality needs to be compatible with the twist. This, in particular, means that the Hodge operator should be star linear, see~\eqref{eq:star-linear}. In \cite{Meier:2023lku}, it was shown that the star linearity of the Hodge operator is satisfied if every element in the support of the twist commutes with the Hodge operator. As discussed previously, for twists including the dilatation generator in its support, this is not satisfied by the standard Hodge operator. Instead, we will consider the modified Hodge operator defined in \eqref{eq:scaleInvHodgeUndeformed}, which by construction commutes with scale transformations. We will now follow the ideas in \cite{Meier:2023lku} in order to find  explicitly the appropriate deformed Hodge operator.

\subsection{Deformed Levi-Civita symbol}
As we  mentioned in footnote \ref{foot:constantR}, in our setup the $\mathcal R$-matrix evaluated on basis forms becomes independent of spacetime positions. Hence, it can be factored out of the star wedge product, cf. \eqref{eq:R-antisym}. This allows us to define a generalised, or deformed, Levi-Civita symbol that collects the coefficients of a given top form with respect to a canonical volume form $\dd^4\xi$ as\footnote{There is in fact another natural way to propose a definition for a deformed Levi-Civita symbol, which turns out to differ from the one above. While we will not  use it in the main text of the paper, we refer to appendix \ref{app:alternative-epsilon} for this  alternative Levi-Civita symbol.}
\begin{equation}
\label{eq:defEpsilon}
\varepsilon_\star^{\mu\nu\rho\sigma}\dd^4\xi=\dd \xi^\mu \wedge_\star \dd \xi^\nu \wedge_\star \dd \xi^\rho \wedge_\star \dd \xi^\sigma,
\end{equation}
where $\varepsilon_\star^{\mu\nu\rho\sigma}$ is star commutative.   For  the volume form we will choose the scale invariant top form
\begin{equation}
\dd^4\xi=\dd \xi^0\wedge \dd \xi^1 \wedge \dd \xi^2 \wedge \dd \xi^3=\dd \xi^0\wedge_\star \dd \xi^1 \wedge_\star \dd \xi^2 \wedge_\star \dd \xi^3,
\end{equation}
where the scale invariant basis one-form is defined as
\begin{equation}
    \dd\xi^\mu=H \star \dd x^\mu. 
\end{equation}
Note that, due to the additional non-trivial star product,  this is not the same scale-invariant one-form as defined in the undeformed case in equation \eqref{eq:undefScaleInvOneForm}. For simplicity, we will employ an abuse of notation and continue to denote it by $\dd \xi^\mu$. 

In general, the star-wedge product of scale-invariant basis forms is $\mathcal R$-antisymmetric. Nevertheless, precisely because the one-forms $\dd\xi^\mu$ are scale invariant, when evaluating the $\mathcal R$-matrix the scale transformation part  vanishes, and the $\mathcal R$-matrix reduces to an object living just in the Poincar\'e algebra. When evaluating $\mathcal R$ on scale-invariant basis one-forms $\dd \xi^\mu$, we will then use the new notation $\tensor{\hat{R}}{_\rho^\mu_\sigma^\nu}$ so that\footnote{Note that the $\mathcal R$-matrix acting only on scale-invariant objects effectively reduces to an $\mathcal R$-matrix  living only in the Poincar\'e algebra. The same comment applies to the twist as well. In particular, by putting a hat over the $\mathcal R$-matrix and the twist, we will indicate that they are evaluated in a representation of scaleless objects, such as $\dd \xi^\mu$. Therefore, for a function $f(x)$, we will write the star product $\dd \xi^\mu\star f=\dd \xi^\mu \tensor{\hat{\bar{F}}}{_\nu^\mu}f$, and similarly $\dd \xi^\mu\wedge_\star\dd \xi^\nu=\tensor{\hat{\bar{F}}}{_\rho^\mu_\sigma^\nu}\dd\xi^\rho\wedge\dd\xi^\sigma$.}
\begin{equation}
    \dd\xi^\mu\wedge_\star\dd\xi^\nu=-\tensor{\hat{R}}{_\rho^\mu_\sigma^\nu}\dd\xi^\sigma\wedge_\star\dd\xi^\rho.
\end{equation}
As a consequence, the scale-invariant Levi-Civita symbol $\varepsilon_\star^{\mu\nu\rho\sigma}$ is identical to the one discussed in \cite{Meier:2023kzt}, meaning that only the contributions of Poincar\'e generators will be non-trivial. Importantly, this  scale-invariant Levi-Civita symbol  is graded cyclic,
\begin{equation}\label{eq:grad-cycl-eps-star}
    \varepsilon_\star^{\mu\nu\rho\sigma}=-\varepsilon_\star^{\nu\rho\sigma\mu}.
\end{equation}

\subsection{$\mathcal{R}$-antisymmetric projectors}
In \cite{Meyer:1994wi} an independent construction of deformed Levi-Civita symbols for $q$-Minkowski space has been discussed and has been further generalised to other braided spaces in \cite{Majid:1994mh}. Although, we start our discussion of non-commutative spacetime from the perspective of a twisted setting, which differs from the setting of $q$-Minkowski, most of the general structure discussed in \cite{Meyer:1994wi,Majid:1994mh} applies in our setting as well. In particular, the non-commutativity of spacetime coordinates is captured by an $\mathcal{R}$-matrix, which satisfies the Yang-Baxter equation, and that can be used to define projectors to totally $\mathcal{R}$-antisymmetric tensors. These can be obtained as contractions of the deformed Levi-Civita symbol, similarly to how projectors onto totally antisymmetric tensors arise from contractions of the undeformed Levi-Civita symbol.

Concretely, we define the projectors onto $\mathcal{R}$-antisymmetric tensors as
\begin{equation}
\begin{aligned}
P_0&=1\\
P_1&=1\\
P_k &= \frac{1}{k}P_{(k-1);2\dots k}\left(\mathds{1}+\sum_{i=2}^{k}(-1)^{i}\mathcal{\check{R}}_{12}\mathcal{\check{R}}_{23}\dots \mathcal{\check{R}}_{i-1,i}\right),
\end{aligned}
\end{equation}
where 
\begin{equation}
\label{eq:epContract}
\begin{aligned}
\mathcal{\check{R}}_{i,i+1}(\dd x^{\mu_1}\otimes \dots\otimes \dd x^{\mu_i} \otimes \dd x^{\mu_{i+1}} \otimes \dots \otimes \dd x^{\mu_k})&=\tensor{R}{_{\nu_1}^{\mu_i}_{\nu_2}^{\mu_{i+1}}}\dd x^{\mu_1}\otimes \dots\otimes \dd x^{\nu_2} \otimes \dd x^{\nu_{1}} \otimes \dots \otimes \dd x^{\mu_k}\\
P_{(k-1);2\dots k}(\dd x^{\mu_1}\otimes \dd x^{\mu_{2}}\otimes \dots \otimes \dd x^{\mu_{k}})&=\dd x^{\mu_1}\otimes P_{k-1}(\dd x^{\mu_2}\otimes \dots \otimes \dd x^{\mu_{k}}).
\end{aligned}
\end{equation}
By using the Yang-Baxter equation for the $\mathcal R$-matrix, it is rather straightforward to show that  the latter projectors are idempotent and $\mathcal{R}$-antisymmetric. 
According to \cite{Majid:1994mh}, the latter projectors can also be obtained via contractions of the deformed Levi-Civita symbol as
\begin{equation}
\tensor{\varepsilon}{_\star^{\rho_{n}\dots\rho_{k+1}\mu_1\dots\mu_k}}\tensor{\varepsilon}{^\star_{\rho_{k+1}\dots\rho_n\nu_k\dots\nu_1}}=-(-1)^{\frac{n(n-1)}{2}}k!(n-k)!\tensor{P}{^{\mu_1\dots\mu_k}_{\nu_1\dots\nu_k}}.
\end{equation}
Hence, we can finally express the Levi-Civita symbol in terms of the $\mathcal R$-matrix via the projectors as
\begin{equation}
\begin{aligned}
\varepsilon_\star^{\mu_1\dots \mu_n}\sim \tensor{P}{^{\mu_1\dots\mu_n}_{(n-1)\dots0}}=\tensor{\varepsilon}{^{\mu_1\dots\mu_n}}\tensor{\varepsilon}{_{0\dots(n-1)}},
\end{aligned}
\end{equation}
where the factor of proportionality can depend on the deformation parameter.

\subsection{Deformed Hodge duality}
Having a good definition of a deformed Levi-Civita tensor, we now define the  deformed Hodge dual  of a basis star-form by
\begin{equation}
\begin{aligned}
*\dd \xi^{\mu_1} \wedge_\star \dots \wedge_\star \dd \xi^{\mu_k} &= \frac{(-1)^{\sigma(k)}}{(n-k)!}\tensor{\varepsilon}{^\star_{\mu_{n}\dots\mu_{k+1}}^{\mu_1\dots\mu_k}} \dd \xi^{\mu_{k+1}} \wedge_\star \dots \wedge_\star \dd \xi^{\mu_n},
\end{aligned}
\end{equation}
where $\sigma(k)=\frac{(n-k)(n+k-1)}{2}$. This definition is completely analogous to the Hodge duality defined in~\eqref{eq:scaleInvHodgeUndeformed} in the untwisted setup. Now twisted wedge products are used and, importantly, the Levi-Civita tensor is the deformed one defined in~\eqref{eq:defEpsilon}. It is clear that even in the deformed setup we still have
\begin{equation}
\begin{aligned}
\mathcal{L}_{X}(*\dd \xi^{\mu_1} \wedge_\star \dots \wedge_\star \dd \xi^{\mu_k})&=*\mathcal{L}_{X}(\dd \xi^{\mu_1} \wedge_\star \dots \wedge_\star \dd \xi^{\mu_k})\\
\mathcal{D}(*\dd \xi^{\mu_1} \wedge_\star \dots \wedge_\star \dd \xi^{\mu_k})&=*\left( \mathcal{D}(\dd \xi^{\mu_1} \wedge_\star \dots \wedge_\star \dd \xi^{\mu_k}) \right),
\end{aligned}
\end{equation}
for any element $X$ of the Poincar\'e algebra. 
Hence, the deformed generalised Hodge duality is star linear:
\begin{equation}\label{eq:star-linear}
\begin{aligned}
*(g\star \omega)&=*(\bar f^\alpha g  \bar f_\alpha\omega)=\bar f^\alpha g \bar f_\alpha(*\omega)=g\star *\omega,
\end{aligned}
\end{equation}
and similarly $*(\omega \star g)=(*\omega) \star g$.
Furthermore, \eqref{eq:epContract} ensures that the deformed Hodge operator defines an actual duality on basis star forms as
\begin{equation}
\begin{aligned}
**\dd \xi^{\mu_1} \wedge_\star \dots \wedge_\star \dd \xi^{\mu_k} &= \frac{(-1)^{\sigma(k)}}{(n-k)!}\tensor{\varepsilon}{^\star_{\mu_{n}\dots\mu_{k+1}}^{\mu_1\dots\mu_k}} *(\dd \xi^{\mu_{k+1}} \wedge_\star \dots \wedge_\star \dd \xi^{\mu_n})\\
&= \frac{(-1)^{\sigma(k)+\sigma(n-k)}}{(n-k)!k!}\tensor{\varepsilon}{_\star^{\mu_1\dots\mu_k}_{\mu_{n}\dots\mu_{k+1}}}\tensor{\varepsilon}{_\star^{\mu_{k+1}\dots\mu_n}_{\nu_{k}\dots\nu_{1}}} \dd \xi^{\nu_{1}} \wedge_\star \dots \wedge_\star \dd \xi^{\nu_k}\\
&=(-1)^{k(n-k)+\frac{n(n-1)}{2}}\frac{1}{(n-k)!k!}\tensor{\varepsilon}{_\star^{\mu_1\dots\mu_k}_{\mu_{n}\dots\mu_{k+1}}}\tensor{\varepsilon}{^\star_{\nu_{k}\dots\nu_{1}}^{\mu_{k+1}\dots\mu_n}} \dd \xi^{\nu_{1}} \wedge_\star \dots \wedge_\star \dd \xi^{\nu_k}\\
&=(-1)^{k(n-k)}\dd \xi^{\mu_1} \wedge_\star \dots \wedge_\star \dd \xi^{\mu_k}.
\end{aligned}
\end{equation}
Together with the star linearity, this carries over to an arbitrary $k$-form $\omega$ as well:
\begin{equation}
\begin{aligned}
**\omega &= (-1)^{k(n-k)} \omega.
\end{aligned}
\end{equation}
Importantly, using the contraction identities between deformed Levi-Civita symbols and the commutation relations of forms and functions, we find
\begin{equation}
\begin{aligned}
&\dd \xi^{\mu_1} \wedge_\star \dots \wedge_\star \dd \xi^{\mu_k} \wedge_\star *(\dd \xi^{\nu_1} \wedge_\star \dots \wedge_\star \dd \xi^{\nu_k})\\
=&\frac{(-1)^{\sigma(k)}}{(n-k)!}\tensor{\varepsilon}{^\star_{\rho_{n-k}\dots\rho_1}^{\nu_1\dots\nu_k}}\varepsilon_\star^{\mu_1\dots\mu_k\rho_1\dots\rho_{n-k}}\dd^n \xi\\
=&\frac{(-1)^{k(k-n)+\sigma(k)}}{(n-k)!} \tensor{\varepsilon}{^\star_{\rho_1\dots\rho_{n-k}}^{\mu_1\dots\mu_k}}\varepsilon_\star^{\rho_{n-k}\dots\rho_1\nu_1\dots\nu_k}\dd^n \xi\\
=&(-1)^{k(n-k)}*(\dd \xi^{\mu_1} \wedge_\star \dots \wedge_\star \dd \xi^{\mu_k}) \wedge_\star \dd \xi^{\nu_1} \wedge_\star \dots \wedge_\star \dd \xi^{\nu_k}.
\end{aligned}
\end{equation}
It follows that for arbitrary $k$-forms $\omega$ and $\chi$
\begin{equation}
\begin{aligned}
\omega \wedge_\star *\chi &= \omega^\star_{\mu_1\dots \mu_k} \star \dd\xi^{\mu_1}\wedge_\star \dots \wedge_\star \dd\xi^{\mu_k}\star \chi^\star_{\nu_1\dots\nu_k}\wedge_\star*\left( \dd\xi^{\nu_1} \wedge_\star\dots\wedge_\star \dd \xi^{\nu_k} \right)\\
&=(-1)^{k(n-k)}\omega^\star_{\mu_1\dots \mu_k} \star \tensor{\hat{R}}{_{\rho_1}^{\mu_1}}\dots\tensor{\hat{R}}{_{\rho_k}^{\mu_k}}\chi^\star_{\nu_1\dots\nu_k}\star*\left(\dd\xi^{\rho_1}\wedge_\star \dots \wedge_\star \dd\xi^{\rho_k}\right) \wedge_\star \dd\xi^{\nu_1} \wedge_\star\dots\wedge_\star \dd \xi^{\nu_k}\\
&=(-1)^{k(n-k)}(*\omega)\wedge_\star\chi.
\end{aligned}
\end{equation}
For star products that are graded-cyclic under integrations, this is sufficient to define a symmetric inner product of forms as
\begin{equation}
\begin{aligned}
\Braket{\omega,\chi}=\int \omega \wedge_\star *\chi=(-1)^{k(n-k)}\int *\omega \wedge_\star \chi = \int \chi \wedge_\star *\omega = \Braket{\chi,\omega}.
\end{aligned}
\end{equation}

The deformed Hodge operator will be useful to define further structures such as the Laplacian and the codifferential. Both will play a role in finding the deformed equations of motions for field theories built on the  non-commutative structures.

\subsection{The coderivative and the Laplacian}\label{sec:Laplacian}
In a given field theory, the equations of motions for a given field usually include the Laplacian $\Delta$ of the underlying spacetime  acting on the field. It can be defined in the usual way when acting on a differential form $\omega$ as
\begin{equation}
\begin{aligned}
\Delta \omega &= \delta \dd \omega + \dd \delta \omega,
\end{aligned}
\end{equation}
where $\delta$ denotes the codifferential defined by
\begin{equation}
\begin{aligned}
\delta = *\dd *.
\end{aligned}
\end{equation}
Since both the Hodge operator and the exterior derivative commute with Poincar\'e generators and with scale transformations generated by $\mathcal{D}$, the coderivative commutes with them as well.
Using the generalised undeformed Hodge operator, this results in a Laplacian acting on a function $f$ as
\begin{equation}
\begin{aligned}
\Delta f=\partial_\mu\partial^\mu f H^{-2}.
\end{aligned}
\end{equation}
Similarly, on a $k$ form $\omega$, the Laplacian becomes
\begin{equation}
\begin{aligned}
\Delta \omega = \partial_\mu\partial^\mu\omega_{\mu_1\dots\mu_k}\dd x^{\mu_1}\wedge\dots\wedge \dd x^{\mu_k} H^{-2}.
\end{aligned}
\end{equation}
Since the exterior derivative and the coderivative commute with Poincar\'e and scale transformations, the Laplacian does as well.

In the deformed setup, the Laplacian on a function becomes
\begin{equation}
\begin{aligned}
\Delta^\star(f)&=*\dd *\dd f\\
&=\partial^\star_\nu\partial^\star_\mu f\star H^{-1} \star *(\dd \xi^\nu\wedge_\star H^{-1} *\dd \xi^\mu)\\
&=\tensor{\bar{\tilde{R}}}{_\rho^{\nu}}\partial^\star_\nu\partial^\star_\mu f\star H^{-2}\star*(\dd\xi^\rho\wedge_\star*\dd\xi^\mu)\\
&=\tensor{\bar{\tilde{R}}}{^{\nu\mu}}\partial^\star_\nu\partial^\star_\mu f\star H^{-2},
\end{aligned}
\end{equation}
where we mainly applied the star linearity of the deformed Hodge operator. Here $\tensor{\bar{\tilde{R}}}{_\rho^{\nu}}$ is the operator $\bar{\tilde{R}}$ evaluated on a basis one-form, and hence it is a matrix. Alternatively, by applying linearity of the Hodge operator, the latter is equivalent to
\begin{equation}
\begin{aligned}
\Delta^\star(f)&=\partial_\nu \partial_\mu f *(\dd x^\nu\wedge*\dd x^\mu)\\
&=\partial_\nu \partial_\mu f H^{-2}*\left((H\dd x^\nu)\wedge*(H\dd x^\mu)\right)\\
&=\partial_\nu \partial_\mu f H^{-2}\tensor{\left.\tilde{F}_{op}\right.}{_\rho^\nu}\tensor{\left.\tilde{F}_{op}\right.}{_\sigma^\mu}*(\dd \xi^\rho\wedge*\dd \xi^\sigma)\\
&=\partial_\nu \partial_\mu f H^{-2}\tensor{\left.\tilde{F}_{op}\right.}{_\rho^\nu}\tensor{\left.\tilde{F}_{op}\right.}{_\sigma^\mu}\tensor{\hat{F}}{_{\rho'}^\rho_{\sigma'}^\sigma}*(\dd \xi^{\rho'}\wedge_\star*\dd \xi^{\sigma'})\\
&=\partial_\nu \partial_\mu f H^{-2}\tensor{\left.\tilde{F}_{op}\right.}{_\rho^\nu}\tensor{\left.\tilde{F}_{op}\right.}{_\sigma^\mu}\tensor{\hat{F}}{_{\rho'}^\rho_{\sigma'}^\sigma}\eta^{\rho'\sigma'}\\
&=\partial_\nu \partial_\mu f H^{-2}\eta^{\mu\nu}\\
&=\Delta(f),
\end{aligned}
\end{equation}
where $\tensor{\left.\tilde{F}_{op}\right.}{_\sigma^\mu}$ is defined via $\tensor{\left.\tilde{F}_{op}\right.}{_\sigma^\mu} H=\tensor{\left.F_{op}\right.}{_\sigma^\mu}(H)$.
Note that $\tensor{\left.\tilde{F}_{op}\right.}{_\nu^\mu}$ is a matrix corresponding to a Poincar\'e transformation and hence $\tensor{\tilde{F}}{_\nu^\mu}\tensor{\bar{\tilde{F}}}{_\rho^\sigma}\eta^{\nu\rho}=\eta^{\mu\sigma}$ and $\tensor{\bar{\hat{F}}}{_{\rho'}^\rho_{\sigma'}^\sigma}\eta^{\rho'\sigma'}=\eta^{\rho\sigma}$. To conclude, the important message is that the Laplacian remains undeformed. Similarly, the Laplacian on an arbitrary differential form acts as
\begin{equation}
\begin{aligned}
\Delta^\star(\omega)&=\tensor{\tilde{R}}{^{\mu\nu}}\partial^\star_\mu\partial^\star_\nu \omega^\star_{\rho_1\dots\rho_k}\star H^{-2} \star \dd x^{\rho_1}\wedge_\star \dots\wedge_\star \dd x^{\rho_k}\\
&=\Delta(\omega^\star_{\rho_1\dots\rho_k})\star\dd x^{\rho_1}\wedge_\star \dots\wedge_\star \dd x^{\rho_k}.
\end{aligned}
\end{equation}
Note that the undeformed Laplacian is Lorentz and scale invariant and hence commutes with star products. Thus,
\begin{equation}
\begin{aligned}
\Delta^\star(\omega)=\Delta(\omega)
\end{aligned}
\end{equation}
for any arbitrary differential form $\omega$.

\section{Non-commutative gauge theories}\label{sec:gauge-theories}

Thanks to the twisted differential calculus defined in the previous section, including a suitable Hodge operator, we can proceed writing down gauge-invariant actions for gauge theories, including matter fields. Again, our approach will generalise the construction done for Poincar\'e twists discussed in \cite{Meier:2023kzt,Meier:2023lku} to include scaling transformations in the twist.

In the non-commutative setting, ordinary gauge symmetries are replaced by so-called ``star gauge symmetries''. Starting, for example, from a fundamental scalar field, i.e.~a scalar field in the fundamental representation of the gauge group, its star gauge transformation under a gauge algebra $\mathfrak{g}$ is given by
\begin{equation}
    \delta_\varepsilon\phi(x)=i\varepsilon(x)\star\phi(x).
\end{equation}
From the point of view of the twisting of the Hopf algebra structure of spacetime symmetries, this is the natural generalisation of the gauge symmetry to the deformed setup, since the gauge transformations depend on the spacetime position and hence are affected by the twist.\footnote{As a consequence, we require the gauge algebra to close under the star commutator instead of the normal commutator. As a result, in general, the gauge algebra is enlarged to its universal enveloping algebra, formally generating additional degrees of freedom, cf. \cite{Szabo:2001kg,Jurco:2001rq}. However, there is a non-local map (the Seiberg-Witten map) from star gauge fields to normal gauge fields that implies the absence of the additional degrees of freedom \cite{Seiberg:1999vs}. Recently, there has been a different approach to define so-called braided non-commutative gauge theories \cite{Ciric:2021rhi,Giotopoulos:2021ieg}, which leads to closing gauge algebra without extending to the universal enveloping algebra. However, its relation to string theory is unclear.}

\paragraph{Covariant derivative} In general, the star product appearing in the gauge transformations may  lead to difficulties in covariantizing the partial derivative \cite{Dimitrijevic:2011jg}, as this now acts via the twisted coproduct. As we will show also in section~\ref{sec:pedestrian}, this can make it difficult to define a gauge field transforming in the right way to cancel the non-covariant terms, and when this happens it is common to use the strategy of constructing a covariant exterior derivative. In fact, since the exterior derivative commutes by construction with Lie derivatives (including $\mathcal{D}$), it follows that it acts via the usual Leibniz rule on star products of differential forms.  Therefore, we can conclude that the gauge transformation of the exterior derivative of the field $\phi$ is just
\begin{equation}
    \delta_\varepsilon\dd\phi=i\varepsilon\star\dd\phi+i\dd \varepsilon\star\phi.
\end{equation}
This is the key point that allows for the straightforward generalisation of the construction of the gauge theories in the deformed setting.
In fact, we can now add a gauge one-form $A$, which transforms as
\begin{equation}\label{eq:gauge-tr-A}
    \delta_\varepsilon A=\dd\varepsilon+i\scom{\varepsilon}{A},
\end{equation}
and this is enough to conclude that the covariant derivative
\begin{equation}
    \begin{aligned}
        \mathrm{D}\phi=\dd \phi-i A\star\phi,
    \end{aligned}
\end{equation}
transforms covariantly under gauge transformations. Having presented the main idea of the construction in the case of a fundamental scalar, we now proceed to systematically list the possible gauge covariant/invariant operators that can be constructed.

\subsection{Yang-Mills theory}
Starting from the gauge one-form $A$, transforming as in~\eqref{eq:gauge-tr-A} under gauge transformation, we can define a field strength two-form
\begin{equation}
    G=\dd A-i A\wedge_\star A,
\end{equation}
that transforms star-covariantly by construction
\begin{equation}
    \delta_\varepsilon(G)=i\scom{\varepsilon}{G}.
\end{equation}
By star-linearity of the Hodge operator, it follows that also the Hodge dual of the two-form field strength transforms covariantly under gauge transformations
\begin{equation}
    \delta_\varepsilon(*G)=i\scom{\varepsilon}{*G}.
\end{equation}
Hence, a star-gauge invariant action is given by
\begin{equation}
    S_{NC-YM}=-\frac{1}{2g_{YM}^2}\int\tr~G\wedge_\star *G,
\end{equation}
which reduces to the standard Yang-Mills action in the undeformed limit.
The invariance of this action under gauge transformations follows directly from the cyclicity of the star product under integration, cf. \eqref{eq:intCyc}. Furthermore, it is equivalent to the action defined in \cite{Meier:2023kzt} if the twist is purely in the Poincar\'e algebra. 

\subsection{Matter fields}
Having a consistent gauge-invariant formulation of non-commutative Yang-Mills theory, we would like to couple matter fields to the gauge field in a gauge invariant way, both in the fundamental and adjoint representations. To include fermionic matter, we will use the index-free notation introduced in \cite{Meier:2023kzt} that makes use of  spinorial equivalents of basis one-forms, and that have been referred to as ``half-forms''.
\subsubsection{Scalar fields}
\paragraph{Fundamental fields} In the case of fundamental scalar fields, we will treat the scaleless field $\varphi$ as the fundamental field under gauge transformations. As in the discussion valid for the forms in the twisted differential calculus, in the twisted setup one may define $\varphi$ in two equivalent ways, namely $\varphi=\phi H^{-1}=\phi_\star\star H^{-1}$. The two are related as $\phi=\tilde{\bar{\mathcal F}}(\phi_\star)$, where $\tilde{\bar{\mathcal F}}$ is obtained by evaluating $\bar{\mathcal F}$ on $H^{-1}$ in the second leg.\footnote{Notice that $H^{-1}$ (the inverse of $H$) is unambiguous also in the deformed setting because $H$ transforms non-trivially only under $\mathcal D$, so that $H^{-1}\star H=H\star H^{-1}=HH^{-1}=1$. } We will define gauge transformations of this object as
\begin{equation}
    \delta_\varepsilon\varphi=i\varepsilon\star\varphi,\qquad\qquad
    \delta_\varepsilon\varphi^\dagger=-i\varphi^\dagger\star\varepsilon,
\end{equation}
where we are taking $H$ to be real. Thanks to the star linearity of the Hodge operator, we can construct a gauge invariant kinetic action  as
\begin{equation}
\label{eq:kinPhi}
    S_{NC-\phi}=\int \mathrm{D}\varphi^\dagger\wedge_\star*D\varphi,
\end{equation}
where
\begin{equation}
\begin{aligned}
    \mathrm{D}\varphi&=\dd\varphi-iA\star\varphi &\qquad \mathrm{D}\varphi^\dagger=\mathrm{d}\varphi^\dagger+i\varphi^\dagger\star A.
\end{aligned}
\end{equation}
As the product $\varphi^\dagger\star\varphi$ is gauge invariant, we can formally write down gauge-invariant terms with any even power $2m$ of the field
\begin{equation}
    \int*\left(\left(\varphi^\dagger\star\varphi\right)^{\star m}\right).
\end{equation}
Importantly, when rewriting $\varphi$ in terms of $\phi$, only the scale-invariant terms will show an absence of unwanted powers of $H$. In this sense, only deformations of scale-invariant terms make sense, because otherwise unphysical powers of $H$ would remain. Nevertheless, one could avoid this issue by redefining also the dimensionful parameters in the action. For example, in order to construct a meaningful mass term that is overall independent of $H$, we can rescale the mass parameter. In the absence of the deformation, for example, we can rewrite the mass term as
\begin{equation}
    S_m=m^2\int\dd^4x~\phi^\dagger\phi=H^{-2}m^2\int *(\phi^\dagger H^{-1}\phi H^{-1})=\mu^2\int *(\varphi^\dagger\varphi),
\end{equation}
where $\mu=mH^{-1}$ is the rescaled mass. Note that, due to the dependence on $H$, the mass $\mu$ is charged under $\mathcal{D}$. When switching on the deformation we can then write
\begin{equation}
    S_{NC-m}=\int \mu^2\star*(\varphi^\dagger\star\varphi)=\int*(\mu^2\star\varphi^\dagger \star \varphi)=\mu^2 \int *(\varphi^\dagger \star \varphi).
\end{equation}
In the last step, we have used the cyclicity of the star product under integration to remove one star product. Importantly, the cyclicity is a consequence of the fact that the above term has zero net power of $H$. Therefore, the remaining star product in the last term above cannot be removed, and we conclude that the mass term is genuinely deformed. This situation is different compared to the case of Poincar\'e twists, where the mass term remains undeformed.

Beyond the free theory, we can define interaction terms in a similar way as for the mass term
\begin{equation}
    \int*\left(\gamma_{2m,n}\star\left(\varphi^\dagger\star\varphi\right)^{\star m}\right),
\end{equation}
which is gauge invariant in the twisted setup for a rescaled coupling constant $\gamma_m=g_{2m,n}H^{2m-n}$, where $n$ is the dimension of the spacetime. In four dimensions, only the quartic interaction is scale invariant and therefore will have the standard coupling constant $\gamma_{4,4}=g_{4,4}$ 
\begin{equation}
    S_{\phi^4}=g_{4,4}\int*(\varphi^\dagger \star \varphi\star\varphi^\dagger \star \varphi).
\end{equation}

\paragraph{Adjoint fields} Let us now consider real scalar fields transforming in the adjoint representation of the gauge group. We will again treat the scaleless field as fundamental, so that we declare that its gauge transformation is
\begin{equation}
    \delta_\varepsilon\varphi=i\scom{\varepsilon}{\varphi}.
\end{equation}

We can therefore take the covariant derivative to be
\begin{equation}
    \mathrm{D}\varphi=\dd\varphi-i\scom{A}{\varphi},
\end{equation} 
and a consistent kinetic action is then given by
\begin{equation}
    S_{NC-\phi}=\int\tr~\mathrm{D}\varphi\wedge_\star*\mathrm{D}\varphi,
\end{equation}
which is gauge invariant because of the star linearity of the Hodge operator and integral cyclicity. Going beyond the kinetic action, we would like to generate interactions and possible mass terms. However,  a consistent interaction term can only be defined for scale-invariant interactions, which excludes also mass terms. The allowed interactions are only of the form
\begin{equation}
    S_{NC-int}=\int\tr~ *(\varphi^{\star 4}).
\end{equation}
Mass terms and other types of
interaction terms are excluded because, due to the lack of integral cyclicity for non-scale-invariant terms, they would not be gauge invariant. It is curious that in the case of fundamental scalars, instead, these terms are in principle possible.

\subsubsection{Fermionic fields}
To include fermions coupled to the gauge field, both in the fundamental and adjoint representation, we will generalise the construction in \cite{Meier:2023lku} to include scaling transformations. As in \cite{Meier:2023lku}, we introduce  a basis of left- and right-handed Grassmann-odd basis spinors $s_\alpha$ and $\bar s^{\dot\alpha}$, so that the Lie derivative with respect to a vector field $X$ acts on  them as
\begin{equation}
\begin{aligned}
        \mathcal{L}_Xs_\alpha&=-\frac{1}{8}(\partial_\mu X_\nu-\partial_\nu X_\mu)\sigma^{\mu}_{\alpha\dot\alpha}\sigma^{\nu\dot\alpha\beta}s_\beta,\\
        \mathcal{L}_X\bar s^{\dot\alpha}&=-\frac{1}{8}(\partial_\mu X_\nu-\partial_\nu X_\mu)\sigma^{\mu\dot\alpha\alpha}\sigma^{\nu}_{\alpha\dot\beta}\bar s^{\dot\beta},
\end{aligned}
\end{equation}
cf. appendix \ref{app:BasisSpinors} for more details on basis spinors and how they behave in the deformed setup. In \cite{Meier:2023lku} Weyl fermions were expressed in an index-free notation as $\psi=\psi^\alpha s_\alpha$ and $\bar\psi=\bar\psi_{\dot\alpha} \bar s^{\dot\alpha}$, so that their Poincar\'e transformations are entirely captured by the spinorial Lie derivative defined in \cite{SpinorLie}. However, this is not enough to capture correctly their scaling dimension.
In order to make these fields scaleless, we need to add the right powers of the auxiliary $H$ as already done for the scalar fields. Hence, we define
\begin{equation}
\begin{aligned}
    \Psi&=\psi H^{-\frac{3}{2}}&\qquad\bar\Psi&=\bar\psi H^{-\frac{3}{2}}.
\end{aligned}
\end{equation}
In the following, we will treat these scaleless fields as the fundamental objects and  assign gauge transformation rules to them. Furthermore, we define a  scaleless one-form $\sigma$ that is Grassmann even and that is built out of the Pauli matrices $\sigma^\mu_{\alpha\dot\alpha}$ as
\begin{equation}
    \sigma=\sigma_{\mu\alpha\dot\alpha}s^\alpha \bar s^{\dot\alpha}dx^\mu H.
\end{equation}
Notice that this is not exactly the $\sigma$ defined in~\cite{Meier:2023lku}, because of the extra $H$ in the definition. We can understand this as the one-form associated with the Pauli matrices that is both   scale- and Poincar\'e-invariant. It will then be natural to use this object to construct  the kinetic action for our fermionic fields.

\paragraph{Fundamental fields} In the case of fundamental fermions, the scaleless Weyl fermions will transform as
\begin{equation}
\begin{aligned}
    \delta_\varepsilon\Psi&=i\varepsilon\star\Psi &\qquad \delta_\varepsilon\bar\Psi&=-i\bar\Psi\star\varepsilon.
\end{aligned}
\end{equation}
As the scale-invariant Pauli one-form $\sigma$ is star-commutative, a gauge-invariant kinetic action is given by
\begin{equation}
    S_{NC-\psi}=\int\dd^2s\dd^2\bar s\int \bar\Psi*\sigma\wedge_\star*\mathrm{D}\Psi,
\end{equation}
where the covariant derivative acts as
\begin{equation}
    \mathrm{D}\Psi = \dd \Psi - iA\star\Psi,\qquad\qquad
    \mathrm{D}\bar\Psi = \dd \bar\Psi + i\bar\Psi\star A.
\end{equation}
In this case the gauge invariance of the action follows simply from the star-linearity of the Hodge operator, and one does not need to use cyclicity of the star product.
A Dirac fermion is built out of a left-handed and a right-handed Weyl-spinors $\Psi$ and $\bar \Xi$ respectively, where both of them transform in the fundamental representation of the gauge group. Hence, we can form Yukawa interactions of the form
\begin{equation}
    S_{NC-Yukawa}=\int*\left(\int \dd^2\bar s~ \bar \Psi\star \varphi^\dagger\star \bar \Xi + \int \dd^2s~ \Xi\star \varphi\star \Psi \right),
\end{equation}
where $\varphi$ is an adjoint scalar field. Note that, as for the scalar fields, a Dirac mass term is allowed for a rescaled mass $\mu=mH^{-1}$ as well.

\paragraph{Adjoint fields} For adjoint fermions, the gauge transformations are given by
\begin{equation}
    \begin{aligned}
        \delta_\varepsilon\Psi&=i \scom{\varepsilon}{\Psi} &\qquad \delta_\varepsilon\bar\Psi&=i \scom{\varepsilon}{\bar\Psi},
    \end{aligned}
\end{equation}
and a kinetic action  is given by
\begin{equation}
    S_{NC-\psi}=\int\dd^2s\dd^2\bar s\int\tr~\bar \Psi\star\sigma\wedge_\star*\mathrm{D}\Psi ,
\end{equation}
where the covariant derivative is now given by
\begin{equation}
    \mathrm{D}\Psi = \dd \Psi - i\scom{A}{\Psi},\qquad\qquad
    \mathrm{D}\bar\Psi = \dd \bar\Psi + i\scom{\bar\Psi}{ A}.
\end{equation}

In this case, gauge invariance of the action follows  from integral cyclicity and star linearity of the Hodge operator, as in the case of adjoint scalars.
In the presence of  an adjoint scalar field, we can define again a deformed Yukawa interaction as
\begin{equation}
    S_{NC-Yukawa}=\int*\tr~\left(\int \dd^2\bar s~ \bar \Psi\star  \varphi^\dagger\star \bar \Xi + \int \dd^2s~ \Xi\star \varphi\star \Psi \right).
\end{equation}
Mass terms are again forbidden by the lack of gauge invariance and non-vanishing dependence on $H$ as for adjoint scalar fields.

\section{An alternative approach}\label{sec:pedestrian}

Compared to the previous sections, in this section we consider an alternative approach. In particular, rather than developing and employing a twisted differential calculus, we work directly with fields and their derivatives. We will show how gauge covariance/invariance can be a guiding principle to define star products that can be used to construct twist non-commutative gauge theories.

The strategy will be to start with a star product implementing a twist that, as in the previous sections, is constructed out of generators acting as Lie derivatives. For pedagogical reasons, we will consider the case of a scalar field $\phi$ transforming in the adjoint representation of the gauge group and try to couple it to a gauge field when switching on the twist. Looking for a good definition of $D_\mu\phi$ (namely the covariant derivative of $\phi$), we will see that a deformation of its naive expression is needed if we want $D_\mu\phi$ to transform covariantly under gauge transformations.

The underlying reason why a modification of the definition of $D_\mu\phi$ is necessary is that in general partial derivatives $\partial_\mu$ do not follow the Leibniz rule for the star product $\star$  constructed with the twist. In the undeformed setup, if the infinitesimal gauge transformation of the field is $\delta_g\phi= i\left[ \epsilon , \phi\right]$, then the gauge transformation of the derivative of the field is $\delta_g(\partial_\mu \phi)=\partial_\mu(\delta_g\phi)= i\left[ \epsilon , \partial_\mu\phi\right]+i\left[ \partial_\mu\epsilon , \phi\right]$. Importantly, the Leibniz rule is a crucial ingredient to get to this formula. This, together with the gauge transformation for the vector field $\delta_g A_\mu= i\left[ \epsilon ,A_\mu\right]+\partial_\mu\epsilon$, is enough to conclude that $D_\mu\phi=\partial_\mu\phi-i\left[ A_\mu , \phi\right]$ transforms covariantly under gauge transformation, i.e. $\delta_g D_\mu\phi= i\left[ \epsilon , D_\mu\phi\right]$. When the Leibniz rule does not hold, the gauge transformation of $\partial_\mu\phi$ looks different, and the definition of the covariant derivative should be modified accordingly.

The fact that in general $\partial_\mu$ does not follow the Leibniz rule is equivalent to the statement that in general translations do not act with a trivial coproduct
\begin{equation}
     \Delta_{\mathcal{F}} (p_\mu)=\mathcal{F}(p_\mu\otimes 1+1\otimes p_\mu)\mathcal{F}^{-1}\neq p_\mu\otimes 1+1\otimes p_\mu.
\end{equation}
If the translations $p_\mu=-i\partial_\mu$ do not commute with  the generators in $\mathcal{F}$, we can expect that, in the presence of the twist, their action on tensor-product spaces is not just the trivial one. A special case, of course, is that of the Groenewold-Moyal deformation, where $\mathcal F$ is constructed with $p_\mu$ only. Because of the commutativity of these generators, the Leibniz rule holds and the usual covariant derivative works. Another example is the dipole deformation of~\cite{Guica:2017mtd} where the twist is constructed with a lightcone momentum and an R-symmetry. However, the issue arising from the non-trivial coproduct of translations appears for example with twists that contain Lorentz generators and that have been considered in~\cite{Aschieri:2006ye,Meier:2023lku,Meier:2023kzt}. There, the problem was solved by constructing a twisted differential calculus, in the spirit of the previous sections. Here we will show how one can get to the same solution in an alternative approach. We will actually start our discussion with twists that were not considered in~\cite{Meier:2023lku}. They are constructed with the dilatation generator $D$, and they allow us to make our point in a simpler setup.

Finally, we will show that the modifications suggested by gauge covariance can be encoded into the definition of a different star product that, to distinguish it from the previous one, we call $\hstar$. In other words, the twist of the gauge theory can be understood just as a replacement of the usual product of fields with the star product $\hstar$. The difference between $\star$ and $\hstar$ is that in the former case the generators appearing in the twist act as Lie derivatives, while in the latter case they implement ``active transformations'' of the fields. 

\subsection{Twists with dilatation and an internal symmetry}\label{sec:D-Q}
For simplicity, we start by considering the twist
\begin{equation}\label{eq:twist-DQ}
    \mathcal{F}=e^{-i\frac{\lambda}{2} D\wedge Q},
\end{equation}
 where $Q$ is the generator of an internal symmetry, i.e.~a symmetry that does not act on spacetime coordinates and only acts on fields. We assume that we work with eigenstates of this symmetry, so that fields have a well-defined charge $Q\phi_a=q_a\, \phi_a$. Importantly, here we take the generator $D$ of scale transformations to act as a Lie derivative, identifying $D=-ix^\mu\partial_\mu$. Knowing that $D$ and $p_\mu$ do not commute, we can anticipate that the Leibniz rule does not hold in this case. In fact, from
\begin{equation}
        \Delta_{\mathcal{F}} (p_\mu)=\mathcal{F}(p_\mu\otimes 1+1\otimes p_\mu)\mathcal{F}^{-1}=p_\mu\otimes e^{\frac{\lambda}{2}Q}+e^{-\frac{\lambda}{2}Q}\otimes p_\mu,
\end{equation}
we may write
\begin{equation}
        \partial_\mu(\phi_a\star\phi_b)=\mu_{\mathcal F}(  \Delta_{\mathcal{F}} (\partial_\mu)(\phi_a,\phi_b))=\partial_\mu\phi_a\star e^{\frac{\lambda}{2}q_b}\phi_b+e^{-\frac{\lambda}{2}q_a}\phi_a\star \partial_\mu\phi_b.
\end{equation}
We can now try to construct a gauge-invariant kinetic term for a complex scalar $\phi$. We start by assuming the gauge transformation for a scalar in the (twisted) adjoint representation
\begin{equation}
    \delta_g\phi=i[\epsilon\stackrel{\star}{,}\phi] = i \left(\epsilon\star \phi-\phi\star\epsilon\right).
\end{equation}
Using the above formulas and assuming that the gauge parameter $\epsilon$ has vanishing $Q$-charge ($q_\epsilon=0$), one can check that the gauge transformation of the derivative of the field is
\begin{equation}
    \delta_g(\partial_\mu\phi)=\partial_\mu(\delta_g\phi)=i\left([\epsilon\stackrel{\star}{,}\partial_\mu\phi]+\zeta\, \partial_\mu\epsilon\star\phi-\zeta^{-1}\, \phi\star\partial_\mu\epsilon\right),
\end{equation}
where we defined
\begin{equation}
    \zeta=e^{\frac{\lambda}{2}q_\phi}.
\end{equation}
We now want to define a covariant derivative for the scalar field. Inspired by the above expression, we take
\begin{equation}
    D_\mu\phi=\partial_\mu\phi-i\left(\ell_1\, A_\mu\star\phi-\ell_2\, \phi\star A_\mu\right).
\end{equation}
One may be tempted to  take $\ell_1=\ell_2$ so that the expression in parenthesis can be written as a star-commutator, but we will see that gauge covariance fixes two different values for the parameters.
To be general, we also allow for a modification of the usual gauge transformation of the gauge field
\begin{equation}
    \delta A_\mu=i\left(\ell_3\, \epsilon\star A_\mu-\ell_4\, A_\mu\star\epsilon\right)+\partial_\mu\epsilon,
\end{equation}
although later we will see that we will need the standard values $\ell_3=\ell_4=1$. Let us now calculate the gauge transformation of $D_\mu\phi$. We find
\begin{equation}\label{eq:cal-gauge-tr-Dphi}
    \begin{aligned}
        \delta D_\mu\phi&=i[\epsilon\stackrel{\star}{,}\partial_\mu\phi]
        +i(\zeta-\ell_1)\partial_\mu\epsilon\star\phi
        -i(\zeta^{-1}-\ell_2)\phi\star \partial_\mu\epsilon\\
        &+\ell_1\ell_3\, \epsilon\star A_\mu\star\phi-\ell_1\, A_\mu\star\phi\star\epsilon
        +\ell_2\ell_4\, \phi\star A_\mu\star\epsilon-\ell_2\, \epsilon\star\phi\star A_\mu\\
        &+\ell_1(1-\ell_4)A_\mu\star\epsilon\star\phi+\ell_2(1-\ell_3)\phi\star\epsilon\star A_\mu.
    \end{aligned}
\end{equation}
We see that we can cancel the terms with the derivative acting on the gauge parameter if we take
\begin{equation}
    \ell_1=\zeta,\qquad\qquad\qquad
    \ell_2=\zeta^{-1}.
\end{equation}
As anticipated, in the presence of the deformation they cannot be taken to be 1. More explicitly, we have
\begin{equation}\label{eq:D-mu-phi-star}
    D_\mu\phi=\partial_\mu\phi-i\left(\zeta\, A_\mu\star\phi-\zeta^{-1}\, \phi\star A_\mu\right).
\end{equation}
Importantly, by introducing these factors we are not cancelling the effect of the star product. In fact, for example, we have
\begin{equation}
  \zeta\, A_\mu(x)\star\phi(x)=\zeta\, A_\mu(\zeta\, x)\phi(x),\qquad\qquad
   \zeta^{-1}\, \phi(x)\star A_\mu(x)=\zeta^{-1}\, \phi(x)A_\mu(\zeta^{-1}\, x),
\end{equation}
(where we used $q_A=0$) so that it is still a non-local product for $\lambda\neq 0$.

To continue, we can cancel the last line in~\eqref{eq:cal-gauge-tr-Dphi} (where $\epsilon$ is in the middle position of the star products) if we take 
\begin{equation}
    \ell_3=\ell_4=1.
\end{equation}
This means that the gauge transformation of the gauge field is the obvious generalisation of the untwisted  one
\begin{equation}\label{eq:gauge-tr-Amu}
    \delta A_\mu=i[\epsilon\stackrel{\star}{,}A_\mu]+\partial_\mu\epsilon.
\end{equation}
Importantly, one can check that with these choices the gauge transformation of the covariant derivative is indeed covariant
\begin{equation}
    \delta D_\mu\phi=i[\epsilon\stackrel{\star}{,} D_\mu\phi].
\end{equation}
Similar considerations apply for the complex conjugate field $\bar\phi$. In that case, taking into account that $q_{\bar\phi}=-q_{\phi}$, one simply has to send $\zeta\to \zeta^{-1}$ everywhere.

To conclude, one can construct a gauge-invariant kinetic term for the complex scalar field by taking
\begin{equation}
    \tr\left(D_\mu\bar\phi\star D^\mu\phi\right),
\end{equation}
where the gauge invariance follows from cyclicity of the trace. In fact, the infinitesimal gauge transformation of the kinetic term is of the form
\begin{equation}
    \tr\left(\epsilon\star D_\mu\bar\phi\star D^\mu\phi-D_\mu\bar\phi\star D^\mu\phi\star\epsilon\right)=\tr\left(\epsilon( D_\mu\bar\phi\star D^\mu\phi)-(D_\mu\bar\phi\star D^\mu\phi)\epsilon\right)=0,
\end{equation}
where some star products are replaced with ordinary products thanks to the fact that both $\epsilon$ and $D_\mu\bar\phi\star D^\mu\phi$ have vanishing $Q$-charge. That allows us to use cyclicity in the last step.

\vspace{12pt}

Interestingly, it is possible to rephrase the above construction into an alternative star product. This is achieved by constructing a twist that, instead of the dilatation generator $ D=-ix^\mu\partial_\mu$ acting as a Lie derivative, has the dilatation generator that includes the extra contribution measuring also the scaling dimension of the fields
\begin{equation}
     D\quad\to\quad -\hat D\equiv D-i\, \sum_I \Delta_I\Phi_I\frac{\delta}{\delta\Phi_I}.
\end{equation}
The reason for the change of sign when going from $D$ to $\hat D$ has to do with conventions that we take for active and passive transformations, see appendix~\ref{app:conv}.
Here $\Phi_I$ is a generic field of the theory and $\Delta_I$ is its scaling dimension. For example, a scalar field in 4 dimensions has $\Delta=1$. In the set $\{\Phi_I\}_I$ one should include together with the fields also their derivatives, with the rule that $\Delta_{\partial 
\Phi_I}=\Delta_{\Phi_I}+1$.
At this point we can consider the alternative twist $\hat{\mathcal F}$ that is obtained from $\mathcal{F}$ simply by replacing $ D$ with $-\hat D$
\begin{equation}
    \mathcal{F}\quad\to\quad\hat{\mathcal{F}}=e^{i\frac{\lambda}{2}\hat D\wedge Q}.
\end{equation}
This defines the new star product $\hstar$
\begin{equation}
    \phi_a\hstar\phi_b=\mu\left(\hat{\mathcal{F}}^{-1}(\phi_a,\phi_b)\right),
\end{equation}
that may be related to the  previous star product $\star$ as
\begin{equation}\label{eq:hstar-phia-phaib}
    \begin{aligned}
        \phi_a\hstar\phi_b&=\mu\left(e^{-i\frac{\lambda}{2}\hat D\wedge Q}(\phi_a,\phi_b)\right)
        =e^{\frac{\lambda}{2}(\Delta_aq_b-q_a\Delta_b)}\mu\left(e^{i\frac{\lambda}{2}D\wedge Q}(\phi_a,\phi_b)\right)\\
        &=e^{\frac{\lambda}{2}(\Delta_aq_b-q_a\Delta_b)}\phi_a{\star}\phi_b.
    \end{aligned}
\end{equation}
Going back to our example of the complex scalar field in the adjoint representation of the gauge group coupled to the vector field, if we take into account that
\begin{equation}
    \Delta_\phi=\Delta_{\bar\phi}=\Delta_A=1,\qquad
    \Delta_\epsilon=0,\qquad
    q_\phi=-q_{\bar\phi},\qquad
    q_A=q_\epsilon=0,
\end{equation}
one finds 
\begin{equation}\label{eq:Aphi-Aeps}
    A_\mu\hstar\phi=\zeta\, A_\mu\star\phi,\qquad\phi\hstar A_\mu=\zeta^{-1}\, \phi\star A_\mu,\qquad
    \epsilon\hstar A_\mu=\epsilon\star A_\mu,\qquad A_\mu\hstar\epsilon= A_\mu\star\epsilon.
\end{equation}
Then all previous formulas written in terms of the star product $\star$ can be rewritten with the new star product $\hstar$ as
\begin{equation}
\begin{aligned}
       & \delta\phi=i[\epsilon\stackrel{\hstar}{,}\phi],\qquad
 &&   \delta\partial_\mu\phi=i\left([\epsilon\stackrel{\hstar}{,}\partial_\mu\phi]+[\partial_\mu\epsilon\stackrel{\hstar}{,}\phi]\right),\\
  &  D_\mu\phi=\partial_\mu\phi-i[A_\mu,\stackrel{\hstar}{,}\phi],\qquad
  &&  \delta D_\mu\phi=i[\epsilon\stackrel{\hstar}{,} D_\mu\phi].
\end{aligned}
\end{equation}
In other words, while it is not true that we can apply the naive rule of replacing the ordinary product of the untwisted gauge theory with the star product $\star$, because doing so we would break gauge covariance/invariance, this naive rule does work with the star product $\hstar$.
Moreover, one may also check that
\begin{equation}
    \tr\left(D_\mu\bar\phi\hstar D^\mu\phi\right)=e^{\frac{\lambda}{2}(4q_\phi)}\tr\left(D_\mu\bar\phi\star D^\mu\phi\right).
\end{equation}
When constructing the kinetic term of the scalar field then, using the star product $\star$ or $\hstar$ gives results that only differ by a normalisation factor, and  gauge invariance is ensured.

\vspace{12pt}

The key point why this is possible is that while $ D$ does not commute with partial derivatives, the operator $\hat D$ does. In fact,
\begin{equation}
\begin{aligned}
    [\hat{D},\partial_\mu]\Phi&=-[ D,\partial_\mu]\Phi+i [\sum_I \Delta_I\Phi_I\frac{\delta}{\delta\Phi_I},\partial_\mu]\Phi\\
    &=-i \partial_\mu\Phi+i((\Delta_\Phi+1)-\Delta_\Phi)\partial_\mu\Phi=0.
\end{aligned}
\end{equation}
By working with $\hat D$, then, we are back to the easy scenario where the generators appearing in the construction of the twist commute with partial derivatives. This implies that the Leibniz rule of derivatives still works, and the covariant derivative can be constructed as in the untwisted case, simply replacing the ordinary product with the star product $\hstar$.

The above discussion is related to the fact that in field theory  one can distinguish between two different types of actions of  symmetries. Supposing that the symmetry acts as $x\to x'$ on coordinates and $\Phi(x)\to\Phi'(x')$ on a generic field, then one can work with ``passive transformations'' where the fields are unchanged and the coordinates transform
\begin{equation}
    \Phi(x')-\Phi(x),
\end{equation}
or with ``active transformations'' where the field transforms but not the coordinates
\begin{equation}
   \hat\delta\Phi\equiv\Phi'(x)-\Phi(x).
\end{equation}
Because, by definition, active transformations do not change spacetime coordinates, it is obvious that they commute with derivatives
\begin{equation}
    [\hat \delta,\partial_\mu]\Phi=0.
    \end{equation}
This fact may be of course verified also using the explicit expressions for active transformations collected in appedix~\ref{app:conv}. The above commutativity relation is then the key for the simplicity of the implementation of the $\hstar$-product that uses active representations of symmetry transformations.

\subsection{Twists with Lorentz and  an internal symmetry}\label{sec:Lor-Q}
Let us now consider the twist
\begin{equation}
    \mathcal{F}=e^{-i\frac{\lambda}{2}\omega^{\mu\nu}M_{\mu\nu}\wedge Q},
\end{equation}
where for the Lorentz generators we take
\begin{equation}
    \begin{aligned}
        M_{\mu\nu}&=2i\, x_{[\mu}\partial_{\nu]}=i(x_\mu\partial_\nu-x_\nu\partial_\mu),\\
        [M_{\mu\nu},M_{\rho\sigma}]&=-i(\eta_{\mu\rho}M_{\nu\sigma}-\eta_{\nu\rho}M_{\mu\sigma}+\eta_{\nu\sigma}M_{\mu\rho}-\eta_{\mu\sigma}M_{\nu\rho}).
    \end{aligned}
\end{equation}
Here $\omega^{\mu\nu}$ are parameters that we may introduce to select the ``polarisation'' of the Lorentz generators used to construct $\mathcal F$.
Now we have
\begin{equation}
    \begin{aligned}
        \mathcal{F}(p_\mu\otimes 1)\bar{\mathcal{F}}&=e^{-i\frac{\lambda}{2}\omega^{\alpha\beta}M_{\alpha\beta}\otimes Q}(p_\mu\otimes 1)e^{i\frac{\lambda}{2}\omega^{\alpha\beta}M_{\alpha\beta}\otimes Q}=p_\nu\otimes F^\nu{}_\mu\\
        \mathcal{F}(1\otimes p_\mu)\bar{\mathcal{F}}&=e^{i\frac{\lambda}{2}\omega^{\alpha\beta}Q\otimes M_{\alpha\beta}}(1\otimes p_\mu)e^{-i\frac{\lambda}{2}\omega^{\alpha\beta}Q\otimes M_{\alpha\beta}}=F_\mu{}^\nu\otimes p_\nu,
    \end{aligned}
\end{equation}
where\footnote{In our conventions, the vectorial representation of the Lorentz transformation $e^{\frac{i}{2}\omega^{\mu\nu}M_{\mu\nu}}\phi(x) =\phi(\Lambda x)$ is $(\Lambda)^\mu{}_\nu=(e^{\omega})^\mu{}_\nu$. For example, on a scalar field we have $e^{\frac{i}{2}\omega^{\mu\nu}M_{\mu\nu}}\phi(x) =\phi(\Lambda x)$.}
\begin{equation}
    F_\mu{}^{\nu}=(\Lambda[\lambda q\omega])_\mu{}^\nu=(\exp(\lambda q\omega))_\mu{}^\nu,
\end{equation}
is a Lorentz transformation whose parameter depends on $\lambda$ and on the $Q$-symmetry value $q$, and we used the property of Lorentz transformations that $F^\mu{}_\nu=(F^{-1})_\nu{}^\mu$.
Therefore
\begin{equation}
    \Delta_{\mathcal F}(p_\mu)=p_\nu\otimes F^\nu{}_\mu+F_\mu{}^\nu\otimes p_\nu.
\end{equation}
It follows again that the partial derivative does not follow the Leibniz rule with the star product $\star$
\begin{equation}\label{eq:dphiaphib}
    \partial_\mu(\phi_a\star\phi_b)=\mu_{\mathcal{F}}(\Delta_{\mathcal F}(\partial_\mu)(\phi_a,\phi_b))=\partial_\nu\phi_a\star F^\nu{}_\mu\phi_b+F_\mu{}^\nu\phi_a\star\partial_\nu\phi_b.
\end{equation}
Nevertheless, also in this case we can introduce a new star product $\hstar$ to achieve $ \partial_\mu(\phi_a\hstar\phi_b)=\partial_\mu\phi_a\hstar \phi_b+\phi_a\hstar\partial_\mu\phi_b$. As earlier, we do that by promoting the Lorentz generators from their Lie-derivative representation to active transformations $M_{\mu\nu}\to - \hat M_{\mu\nu}$. Importantly, the action is different for scalars and vectors
\begin{equation}
    \hat M_{\mu\nu}(\phi(x))=-2ix_{[\mu}\partial_{\nu]}\phi(x),\qquad
    \hat M_{\mu\nu}(\partial_\rho\phi(x))=-2ix_{[\mu}\partial_{\nu]}(\partial_\rho\phi(x))-2i\eta_{\rho[\mu}\partial_{\nu]}\phi(x),
\end{equation}
so that the finite actions are
\begin{equation}
    e^{\frac{i}{2}\omega^{\mu\nu}\hat M_{\mu\nu}}(\phi(x))=\phi(\Lambda^{-1}x),\qquad
    e^{\frac{i}{2}\omega^{\mu\nu}\hat M_{\mu\nu}}(\partial_\rho\phi(x))=\Lambda_\rho{}^\alpha\partial_\alpha\phi(\Lambda^{-1}x).
\end{equation}
If we now define the new twist
\begin{equation}
    \hat{\mathcal{F}}=e^{i\frac{\lambda}{2}\omega^{\mu\nu}\hat M_{\mu\nu}\wedge Q},
\end{equation}
and the corresponding $\hstar$ product, we can easily check that
\begin{equation}
    \phi_a\hstar\phi_b=\phi_a\star\phi_b,\qquad \partial_\mu\phi_a\hstar\phi_b=\partial_\nu\phi_a\star F^\nu{}_\mu\phi_b,\qquad \phi_a\hstar\partial_\mu\phi_b=F_\mu{}^\nu\phi_a\star\partial_\nu\phi_b.
\end{equation}
This is enough to conclude that now
\begin{equation}\label{eq:dhatphiaphib}
    \partial_\mu(\phi_a\hstar\phi_b)=\partial_\mu\phi_a\hstar\phi_b+\phi_a\hstar\partial_\mu\phi_b.
    \end{equation}
This of course was to be expected from the general argument that active transformations commute with partial derivatives.
Once again, we can then define the covariant derivative using the hatted star product, and the gauge invariance of $\tr(D_\mu\bar\phi\hstar D^\mu\phi)$ follows again from $Q_\epsilon=0$ and from the cyclicity of the trace with respect to the standard product.

 \subsection{Twists with Lorentz only}
Let us  consider a twist constructed only out of Lorentz generators. Now we skip the construction of the star product $\star$, and we jump immediately to the construction of the star product $\hstar$ arising form the twist
\begin{equation}
    \hat{\mathcal{F}}=e^{-i\frac{\lambda}{2} \omega^{\alpha\beta}\hat M_{\alpha\beta}\wedge \omega'{}^{\alpha\beta}\hat M_{\alpha\beta}}.
\end{equation}
We do this because we know that the  replacement of the standard product with $\hstar$ gives rise to good covariant derivatives. Up to automorphisms, the only interesting possibility is $\hat{\mathcal{F}}=e^{-i\frac{\lambda}{2} \hat M_{01}\wedge \hat M_{23}}$, and this is in fact the case considered in~\cite{Meier:2023kzt}. It is easy to show that the construction proposed here is equivalent to the one of~\cite{Meier:2023kzt}. In fact,~\cite{Meier:2023kzt} makes use of objects (e.g.~one-forms like $A=dx^\mu A_\mu$) that by construction allow one to replace the hatted star product $\hstar$ with the ``Lie-derivative'' star product $\star$. See section~\ref{sec:equivalence} for more comments on the equivalence of the two formulations.

In this case, the Lagrangian density $\mathscr L$ of the twisted theory is not gauge invariant, but upon integration the action is gauge invariant. In fact, infinitesimally, we have
\begin{equation}
    \int d^dx\ \tr ([\epsilon\stackrel{\hstar}{,}\mathscr L])\sim \int d^dx\ \left(\tr ([\epsilon,\mathscr L])
    -i\lambda\left(\hat M_{01}\epsilon \ \hat M_{23}\mathscr{L}-\hat M_{23}\epsilon \ \hat M_{01}\mathscr{L}\right)+\ldots\right).
\end{equation}
The first term appearing at $\mathcal O(\lambda^0)$  vanishes because of cyclicity of the trace, and the subleading terms at $\mathcal O(\lambda)$ also vanish because they add up to a total derivative. In fact, both $\epsilon$ and $\mathscr{L}$ are scalars so that the expression for $\hat M_{\mu\nu}$ is just the one of $M_{\mu\nu}$, and then for example
 \begin{equation}
     \int d^dx\ \left(x_0x_2\partial_1\epsilon \partial_3\mathscr{L}-x_0x_2\partial_3\epsilon \partial_1\mathscr{L}\right)=0.
 \end{equation}
 Iterating this argument at each order in the $\lambda$-expansion one proves gauge invariance to all orders.

 \subsection{Twists with dilatation and Lorentz}
 We may consider also the twist constructed with the dilatation operator and Lorentz generators as
\begin{equation}
    \hat{\mathcal{F}}=e^{-i\frac{\lambda}{2}\hat D\wedge \omega^{\alpha\beta}\hat M_{\alpha\beta}}.
\end{equation}
As a side remark, we notice that the star product $\star$ defined only through Lie derivatives has complicated rules when considering derivatives of star products of fields. For example, for ${\mathcal{F}}=e^{-i\frac{\lambda}{2}  D\wedge  M_{23}}$ one would have
\begin{equation}
    \begin{aligned}
        \partial_{0,1}(\phi_a\star\phi_b)&=\partial_{0,1}\phi_a\star e^{\frac{\lambda}{2}M_{23}}\phi_b+e^{-\frac{\lambda}{2}M_{23}}\phi_a\star \partial_{0,1}\phi_b,\\
        \partial_2(\phi_a\star\phi_b)&=\partial_2\phi_a\star \cos(\tfrac{\lambda}{2} D)  e^{\frac{\lambda}{2}M_{23}}\phi_b
        +\partial_3\phi_a\star \sin(\tfrac{\lambda}{2} D)  e^{\frac{\lambda}{2}M_{23}}\phi_b\\
        &+\cos(\tfrac{\lambda}{2} D)  e^{-\frac{\lambda}{2}M_{23}}\phi_a\star \partial_2\phi_b
        - \sin(\tfrac{\lambda}{2} D)  e^{-\frac{\lambda}{2}M_{23}}\phi_a\star \partial_3\phi_b,\\
        \partial_3(\phi_a\star\phi_b)&=\partial_3\phi_a\star \cos(\tfrac{\lambda}{2} D)  e^{\frac{\lambda}{2}M_{23}}\phi_b
        -\partial_2\phi_a\star \sin(\tfrac{\lambda}{2} D)  e^{\frac{\lambda}{2}M_{23}}\phi_b\\
        &+\cos(\tfrac{\lambda}{2} D)  e^{-\frac{\lambda}{2}M_{23}}\phi_a\star \partial_3\phi_b
        + \sin(\tfrac{\lambda}{2} D)  e^{-\frac{\lambda}{2}M_{23}}\phi_a\star \partial_2\phi_b.
    \end{aligned}
\end{equation}
While this looks complicated,  we don't need to worry about this, because we  work directly with $\hstar$. Because we can construct a good covariant derivative with $\hstar$, the only task left is to check the gauge invariance of the action. When looking at $\tr ([\epsilon\stackrel{\hstar}{,}\mathscr L])$, then, we can think of  the Lagrangian density as a field with scaling dimension $\Delta_{\mathscr{L}}=d$, where $d$ is  the number of spacetime dimensions. We can do this because of the original assumption that before applying the twist the theory is scale invariant. In formulas, we can then write
\begin{equation}
    \hat{D}\mathscr{L}=i\, x^\mu\partial_\mu \mathscr{L}+i\, \Delta_{\mathscr{L}}\mathscr{L}.
\end{equation}
For consistency,  the gauge parameter $\epsilon$ cannot have a scaling dimension, so we can set $\Delta_\epsilon=0$. As before, the term at $\mathcal O(\lambda^0$) in  $\tr ([\epsilon\stackrel{\hstar}{,}\mathscr L])$ vanishes because of the cyclicity of the trace, so that we are interested in the term at  $\mathcal O(\lambda)$, that in this case reads
\begin{equation}\label{eq:oder-lambda-DM}
\begin{aligned}
    i\lambda\tr\left(\hat D\epsilon\hat M_{23}\mathscr{L}-\hat M_{23}\epsilon\hat D\mathscr{L}\right)&=
    i\lambda\tr[(x^\mu\partial_\mu\epsilon)(x_2\partial_3\mathscr{L}-x_3\partial_2\mathscr{L})\\
    &\qquad -(x_2\partial_3\epsilon-x_3\partial_2\epsilon)(x^\mu\partial_\mu \mathscr{L}+\Delta_{\mathscr{L}}\mathscr{L})].
\end{aligned}
\end{equation}
Assuming the integration over $d^dx$ used to construct the action, we can perform integrations by parts (first for the $\partial_\mu$ coming from $\hat D$ and then for the $\partial_{2,3}$ of $\hat M_{23}$) to relate the first line above to the second line. One finds
\begin{equation}
   i\lambda\tr[(x^\mu\partial_\mu\epsilon)(x_2\partial_3\mathscr{L}-x_3\partial_2\mathscr{L})]\approx
   i\lambda\tr[ (x_2\partial_3\epsilon-x_3\partial_2\epsilon)(x^\mu\partial_\mu \mathscr{L}+d\ \mathscr{L})],
\end{equation}
where we write $\approx$ because the equality is true only up to total derivatives.
We see that the first line of~\eqref{eq:oder-lambda-DM} exactly cancels the second line (up to boundary terms) precisely thanks to $\Delta_{\mathscr{L}}=d$.
Before concluding, let us  once again point out  that the naive construction with the star product $\star$ would not give rise to a gauge-invariant action. In fact, the computation in the case  of $\star$ would be identical to the one performed here, with the exception that we would not have the contributions with $\Delta_{\mathscr L}$. The first and the second line in~\eqref{eq:oder-lambda-DM} would not cancel each other, then, because of an extra term of the form $d\tr[ (x_2\partial_3\epsilon-x_3\partial_2\epsilon)\mathscr{L}]$.

\subsection{Equivalence of the two approaches}\label{sec:equivalence}
It should be clear that the approach of the twisted differential calculus of section~\ref{sec:twist-diff-calc} and the one presented in this section give rise to the same result. To see this more explicitly, we will start by discussing the equivalence in two examples, and then we will provide a general argument.

\subsubsection{Twists with dilatation and an internal symmetry}
Let us take for example the twist considered at the beginning of this section involving the dilatation generator and a charge $Q$. In the spirit of section~\ref{sec:twist-diff-calc}, we want to take a star product implemented by the twist
\begin{equation}
    \mathcal{F}=e^{-i\frac{\lambda}{2} \mathcal D\wedge Q}=e^{-i\frac{\lambda}{2}  D\wedge Q}e^{-i\frac{\lambda}{2} (iH\partial_H\wedge Q)}.
\end{equation}
Here we wrote separately the part with $D$ that acts as a Lie derivative on standard fields (e.g.~$\phi$)  and the part acting on the auxiliary $H$. If we have fields $\phi_a$ of scaling dimension $\Delta_a$, we can construct scaleless fields $\varphi_a=\phi_a H^{-\Delta_a}$, and it is clear that we have
\begin{equation}
    \varphi_a\star \varphi_b = (\phi_aH^{-\Delta_a})\star(\phi_bH^{-\Delta_b})=e^{\frac{\lambda}{2}(\Delta_aq_b-q_a\Delta_b)}(\phi_a\star\phi_b)H^{-2}.
\end{equation}
In the last step, we evaluated explicitly the part of the twist acting on $H$, so that the star product acting on $\phi_a$ and $\phi_b$ is now implemented just by the twist with the Lie derivative as in~\eqref{eq:twist-DQ}. Notice the similarity of this expression with the one in~\eqref{eq:hstar-phia-phaib}.
Applying the same logic, if we now consider $A=A_\mu\dd x^\mu$ we may write
\begin{equation}
\begin{aligned}
    A\star\varphi&=\zeta\,\dd x^\mu H^{-1} A_\mu\star\phi,&\qquad\varphi\star A&=\zeta^{-1}\, \phi\star A_\mu \dd x^\mu H^{-1},\\
    \epsilon\star A&=(\epsilon\star A_\mu  \dd x^\mu),&\qquad A\star\epsilon&=  (\dd x^\mu A_\mu\star\epsilon),
\end{aligned}
\end{equation}
which is the counterpart of~\eqref{eq:Aphi-Aeps}.
This is enough to conclude, for example, that
\begin{equation}
    \delta_\varepsilon A=\dd\varepsilon+i\scom{\varepsilon}{A}=\dd x^\mu(\partial_\mu\varepsilon+i\scom{\varepsilon}{A_\mu}),
\end{equation}
and
\begin{equation}
    \mathrm{D}\varphi=\dd\varphi-i\scom{A}{\varphi}=\dd x^\mu H^{-1}[\partial_\mu\phi-i(\zeta\, A_\mu\star\phi-\zeta^{-1}\, \phi\star A_\mu )],
\end{equation}
which are precisely the gauge transformation in~\eqref{eq:gauge-tr-Amu} and the covariant derivative in~\eqref{eq:D-mu-phi-star}.

To summarise, it is clear that the introduction of the auxiliary $H$ serves precisely the purpose of including the contribution of the scaling dimension of the fields when computing the star product. This is equivalently achieved by promoting the dilatation generator to be represented as an active transformation, as done in section~\ref{sec:D-Q}.

\subsubsection{Twists with Lorentz and an internal symmetry}

To continue with our examples, consider now the twist with a Lorentz transformation and an internal symmetry as in section~\ref{sec:Lor-Q}
\begin{equation}
    \mathcal{F}=e^{-i\frac{\lambda}{2}\omega^{\mu\nu}M_{\mu\nu}\wedge Q}.
\end{equation}
Although this does not involve the dilatation generator, it is instructive to see how the construction of~\cite{Meier:2023lku} is captured by the star product constructed out of active symmetry transformations. In fact, in the spirit of~\cite{Meier:2023lku}, one should define gauge transformations and construct gauge covariant/invariant quantities only when working with forms (or half forms in the case of fermions). Then, instead of $\partial_\mu\phi_a$, for example, one should work with\footnote{In this case it is not necessary to introduce $H$ and $\xi$ because the twist does not involve the dilatation generator.} $\dd \phi_a=\dd x^\mu \partial_\mu\phi_a$. Starting from
\begin{equation}\label{eq:dphi}
    \dd (\phi_a\star \phi_b)=\dd \phi_a\star \phi_b+\phi_a\star \dd \phi_b,
\end{equation}
and using that 
\begin{equation}
     M_{\mu\nu}(\dd x^\rho) = 2i\dd x_{[\mu}\delta_{\nu]}^\rho,
\end{equation}
one can factorise the basis 1-form in~\eqref{eq:dphi} and write
\begin{equation}
    \dd (\phi_a\star \phi_b)=\dd x^\mu\partial_\mu(\phi_a\star \phi_b)=\dd x^\mu\left(\partial_\nu\phi_a\star F^\nu{}_\mu\phi_b+F_\mu{}^\nu\phi_a\star\partial_\nu\phi_b\right).
\end{equation}
This agrees with~\eqref{eq:dphiaphib}, which was already shown to be  equivalent to the expression with the hatted star product of~\eqref{eq:dhatphiaphib}.

\subsubsection{The general argument}

Going beyond examples, the equivalence of the two formulations is simply a consequence of the following simple but general observation. First, consider a field $\Phi$ of scaling dimension $\Delta$. The prescription of section~\ref{sec:twist-diff-calc} is to construct the scaleless field $\Phi H^{-\Delta}$ and then build the twist with $\mathcal D$. However, we can write
\begin{equation}
    \mathcal D(\Phi H^{-\Delta})=(D+iH\partial_H)(\Phi H^{-\Delta})=(D\Phi-i\Delta\Phi)H^{-\Delta}=-(\hat D\Phi)H^{-\Delta},
\end{equation}
so that we can rewrite the action of $\mathcal D$ on the scaleless field as the active version of the symmetry transformation  on the field $\Phi$. Similarly, we may consider the same rewriting for Lorentz transformations. Leaving aside the trivial case of scalars, in the case of  vectors, for example, we have
\begin{equation}
    M_{\mu\nu}(A_\rho\dd x^\rho)=\dd x^{\rho}(M_{\mu\nu}(A_\rho)+2i\eta_{\rho[\mu}A_{\nu]})=-\dd x^\rho\ \hat M_{\mu\nu}(A_\rho),
\end{equation}
so that also in this case the prescription of~\cite{Meier:2023lku} and of section~\ref{sec:twist-diff-calc} is equivalent to considering the active version of the symmetry transformation. This allows to replace the Lie-derivative star product $\star$ by the hatted star product $\hat \star$ for the actions in section \ref{sec:gauge-theories}.

Although naively this may seem enough to prove the equivalence of the two formulations, when considering  the action written with the twisted differential calculus, one should first check that  the basis forms, the basis spinors and the powers of $H$ factor out of the star products,  and combine to the correct index contractions to finally reproduce the action written in the alternative approach.
Importantly, once the hatted star product is used, the basis one forms, the basis spinors and the auxiliary $H$ star-commute. Hence, the hatted star product with these objects reduces to the standard product
\begin{equation}
\begin{aligned}
    \dd x^\mu\wedge_{\hat\star}\dd x^\nu&=\dd x^\mu\wedge\dd x^\nu\\
    f\hat\star\dd x^\mu&=f\dd x^\mu\\
    \omega\hat\star H&=\omega H.
\end{aligned}
\end{equation}
The last point to check is that the Hodge dual reduces to the standard expression with the usual Levi-Civita symbol. In order to see this, consider for example the \emph{deformed}  Hodge operator acting on a basis two-form with normal wedge products. It becomes
\begin{equation}
\begin{aligned}
    *\left( \dd \xi^{\mu_1}\wedge_{\hstar} \dd \xi^{\mu_2} \right)&= *\left( \dd \xi^{\mu_1}\wedge \dd \xi^{\mu_2} \right)\\
    &=\tensor{\hat F}{_{\mu_1'}^{\mu_1}_{\mu_2'}^{\mu_2}}*(\dd \xi^{\mu_1'}\wedge_\star\dd \xi^{\mu_2'})\\
    &=-\frac{1}{2}\tensor{\hat F}{_{\mu_1'}^{\mu_1}_{\mu_2'}^{\mu_2}}\tensor{\left.\varepsilon_\star\right.}{_{\nu_2\nu_1}^{\mu_1'\mu_2'}}\dd \xi^{\nu_1}\wedge_\star \dd \xi^{\nu_2}\\
    &=-\frac{1}{2}\tensor{\hat F}{_{\mu_1'}^{\mu_1}_{\mu_2'}^{\mu_2}}\tensor{\hat {\bar F}}{_{\nu_1'}^{\nu_1}_{\nu_2'}^{\nu_2}}\tensor{\left.\varepsilon_\star\right.}{_{\nu_2\nu_1}^{\mu_1'\mu_2'}}\dd \xi^{\nu_1'}\wedge \dd \xi^{\nu_2'},
\end{aligned}
\end{equation}
where we are using again the notation $\hat F$ to indicate that the twist is acting on scaleless objects.
We stress that the Hodge operator appearing in the starting point of the above equation is the deformed one. At the same time, we can relate the deformed Levi-Civita tensor to the undeformed one as 
\begin{equation}
\begin{aligned}
    \varepsilon_\star^{\mu\nu\rho\sigma}\dd^4\xi&=\dd \xi^\mu\wedge_\star\dd\xi^\nu\wedge_\star\dd\xi^\rho\wedge_\star\dd\xi^\sigma\\
    &=(\dd \xi^\mu\wedge_\star\dd\xi^\nu)\wedge(\dd\xi^\rho\wedge_\star\dd\xi^\sigma)\\
    &=\tensor{\hat {\bar F}}{_{\mu'}^{\mu}_{\nu'}^{\nu}}\tensor{\hat {\bar F}}{_{\rho'}^{\rho}_{\sigma'}^{\sigma}}
    \dd \xi^\mu\wedge\dd\xi^\nu\wedge\dd\xi^\rho\wedge\dd\xi^\sigma\\
    &=\tensor{\hat {\bar F}}{_{\mu'}^{\mu}_{\nu'}^{\nu}}\tensor{\hat {\bar F}}{_{\rho'}^{\rho}_{\sigma'}^{\sigma}}\varepsilon^{\mu'\nu'\rho'\sigma'}\dd^4\xi,
\end{aligned}
\end{equation}
where the cyclicity of the basis top form was used in the second line. Putting the two results together, we can conclude that
\begin{equation}
    *(\dd\xi^\mu\wedge_{\hstar}\dd\xi^\nu)=-\frac{1}{2}\tensor{\varepsilon}{_{\sigma\rho}^{\mu\nu}}\dd\xi^{\rho}\wedge \dd\xi^{\sigma},
\end{equation}
which is the expression for the undeformed Hodge dual of the basis two form. This argument can be similarly applied to all possible basis $k$-forms.

To conclude, as the hatted star product involving $H$ and basis one-forms is trivial, and the Hodge dual of basis objects written in terms of the hatted star product appears to be undeformed as well, the actions constructed with differential forms via the star product $\star$ are identical to those constructed using the alternative approach with the hatted star product $\hat \star$. Importantly, this is true for all terms that are overall independent of $H$ and hence scale invariant.

\section{Twisted symmetry}\label{sec:twist-symm}
In general, one may expect that the deformation breaks at least some of the symmetries of the original action, in particular when the symmetry generators do not commute with the twist. While strictly speaking this is true, it is possible to identify a ``twisted'' realisation of such symmetries, at least at the classical level. The fact that this is possible was known already in the case of the Groenewold-Moyal deformation as shown in~\cite{Chaichian:2004yh,Chaichian:2004za}, and it was then generalised to all twists of the Poincar\'e algebra in~\cite{Meier:2023lku}. 
The identification of the twisted symmetry is in fact natural when working with the star product. While symmetry generators act in the usual way on individual fields, their action on star products of fields need to take into account the non-trivial coproduct. When doing this, one is sure to respect the undeformed commutation relations of the generators, but at the same time a novel (twisted) realisation of the symmetry is achieved.

Here we will generalise the discussion of~\cite{Meier:2023lku}, following then the twisted differential calculus formulation. Importantly, one assumption in our present work is that the undeformed model is not only Poincar\'e invariant but also scale invariant. We will therefore show that there is a way to implement twisted Poincar\'e and scale  transformations.

The main idea is that to define the action of a symmetry generator $X$ on a wedge-star product of fields $\Phi_i$ appearing in a generic term of the Lagrangian, we can do so via the twisted coproduct as
\begin{equation}
    \delta^\star_X(\Phi_1\wedge_\star\Phi_2\wedge_\star\ldots\wedge_\star\Phi_n)=\left.\bigwedge\right._\star(\Delta^{(n)}_{\mathcal{F}}(X)(\Phi_1,\Phi_2,\ldots,\Phi_n)).
\end{equation}
Here $\Phi_i$ denotes a generic field, i.e.~scalars, fermions or vectors, written in the twisted differential calculus formulation, possibly with the exterior derivative or Hodge dual acting on them.  When $X$ is in the Poincar\'e algebra, the reason to have such a definition for the action of the symmetry generator is that---simply as a consequence of the definition of the twist and the associativity of the star product--- the right-hand-side of the above equation is equivalent to the Lie derivative with respect to $X$ acting on the whole star product of fields
\begin{equation}
    \delta^\star_X(\Phi_1\wedge_\star\Phi_2\wedge_\star\ldots\wedge_\star\Phi_n)=\mathcal L_X(\Phi_1\wedge_\star\Phi_2\wedge_\star\ldots\wedge_\star\Phi_n).
\end{equation}
 If $X=D$, the Lie derivative $\mathcal L_D$ is accompanied also by the piece measuring the overall power of $H$. Hence, for all terms with trivial overall power of $H$, the result is again given by the above expression.
The above formula is obvious whenever $\Phi_i$ behaves simply as functions with respect to the symmetry generator $X$, so that the corresponding symmetry action can be implemented via a Lie derivative on each $\Phi_i$. An example of this is when dealing with generators of Poincar\'e acting on scalar fields and, thanks to the introduction of the auxiliary $H$ and the scaleless field $\varphi$, in this paper we have achieved this setup also for the scale transformation $D$. Similarly,  the argument applies also to fermions and vectors if one works with the appropriate objects that transform via Lie derivatives under both Poincar\'e and scale transformations. Finally, as we have argued in section~\ref{sec:twist-diff-calc}, the Lie derivatives by generators of Poincar\'e or by the scale transformations commute with both the exterior derivative and the (deformed) Hodge dual. This implies that, starting from $\delta^\star_X\Phi_i=\mathcal L_X\Phi_i$, the above formula holds for all (kinetic or interaction) terms in the Lagrangian, and whenever $X$ is a generator of the Poincar\'e algebra or the scale transformation.

The important point is that because such Lagrangian terms appear under integration in the action of the theory, the variation of the action is equivalent to the integral of a total derivative, and thus vanishes. This concludes the proof that the deformed action is still invariant under a twisted notion of Poincar\'e and  scale transformations.

Finally, let us mention that we can still talk about a twisted symmetry group of transformations because compositions of the above transformations still close under the usual commutation relations of the symmetry algebra, thanks to the properties of the coproduct. However, its notion on products of fields is twisted as expected.

In this paper we have discussed the notion of twisted symmetries from the point of view of the twisted differential calculus formulation. In~\cite{next-paper} we will do it in the formulation where the twist is constructed out of active transformations.

\section{Planar equivalence theorem}\label{sec:planar}

As the non-commutative deformations change the action of the theory, the Feynman diagrams that appear in perturbative computations of correlation functions and scattering amplitudes will be deformed as well. However, the star product introduces a natural notion of planarity. As the star product is cyclic under integration, the star product in each vertex behaves like single trace interactions in a theory of matrix valued fields. Furthermore, for a gauge theory, the notion of planarity from the star product coincides with the notion of planarity introduced by the rank of the gauge group. This structure has been first used in the case of the Groenewold-Moyal deformation to relate planar diagrams in the deformed theory to their undeformed counterparts \cite{FILK199653}. 
In particular, it was proven that in planar Feynman diagrams the deformed theory exhibits the same divergences as the undeformed one. This is a consequence of the stronger statement that in the planar limit amputated Feynman diagrams (i.e. those where external legs are removed) remain undeformed, so that the only effect of the deformation sits in the twists acting in the external leg propagators.
These facts were generalised in \cite{Meier:2023lku} to all twist-deformations in the Poincar\'e algebra with cyclic star products under integration. In the following, we will extend this planar equivalence theorem to deformations that include the scale transformation in the twist. To prove the theorem, we will have to restrict ourselves to theories that are relativistic and scale invariant. The reason is that one important assumption of the theorem is that the propagators of the twisted theory should remain undeformed. While, in principle, gauge invariance allows us to have mass terms for fundamental matter fields in the twisted theory, these mass terms are affected by the deformation, and then the propagators of massive theories do not remain undeformed.

In the following, we will list the main points needed to prove the planar equivalence theorem, closely following the steps of \cite{Meier:2023lku}. We will not need to give many details because the proof is formally the same as in that paper. In fact, we will simply show that we can fulfil the  preliminary conditions and assumptions of the proof of \cite{Meier:2023lku}, so that the theorem automatically follows. To keep the discussion simple, we will consider the case of Feynman diagrams involving only scalar fields, but the proof can be generalised when including fermions and gauge vectors as explained in \cite{Meier:2023lku}.

\paragraph{Preparing the undeformed diagrams} Before  discussing the deformed Feynman diagrams, one should express the undeformed diagrams in a suitable way. We will denote by $y_k$ the ``external points'' of the diagram, and by $x_k$ the ``internal points'' over which one integrates. The appearance of the internal points can be understood, for example, as coming from the expansion of the exponential of the action inside the path integral description. Typically, we can have powers of fields evaluated at internal points, because we can have quartic interactions for the scalar fields, for example. However, to keep track of the action of the twist, we will employ a sort of ``regularisation'' and evaluate each power of $\Phi$ at a different point $x_k$. Note that $\Phi$ in principle can denote any kind of fields in a given theory. We suppress any kind of index that could appear for readability.

Importantly, we will consider amputated diagrams by stripping off all external propagators.  When considering amputated diagrams, we will leave uncontracted those  fields that are evaluated at internal points $x_k$ and that are supposed to be Wick contracted to fields evaluated at external positions $y_k$. In the context of the planar equivalence theorem, we will refer to these uncontracted fields $\Phi(x_k)$ as ``external'', because in a certain sense they inherit the feature of being external after the amputation. The integrand of a given Feynman diagram is now determined by all these external fields $\Phi(x_k)$ and by the remaining internal propagator structure of the diagram, which we will denote by $I_n$.\footnote{Formally, $I_n$ depends on further internal points, but we will not indicate that in our notation because the only interesting internal points for us are those that are eventually contracted with the external positions.} Hence, up to Wick contractions with the external fields, the integrand of an undeformed Feynman diagram is given by
\begin{equation}
    a_n(I_n;x_1\dots x_n)=I_n\times\prod_{k=1}^n \Phi(x_k).
\end{equation}
\paragraph{The planar equivalence theorem} We say that a deformed diagram corresponds to an undeformed diagram if the undeformed limit of the former gives the latter, i.e. they coincide up to star products. The planar equivalence theorem now states that 
\begin{enumerate}
    \item A deformed planar diagram can be given as the undeformed diagram up to star products  acting exclusively on the external fields:
\begin{equation}
    a^\star_n(I_n;x_1\dots x_n)=I_n\times\underset{k=1}{\overset{n}{\bigstar}}\Phi(x_k).
\end{equation}
\item The remaining star product on the external positions remains cyclic:
\begin{equation}
    I_n\times\underset{k=1}{\overset{n}{\bigstar}}\Phi(x_k)=I_n\times\underset{k=j+1}{\overset{n}{\bigstar}}\Phi(x_k)\star\underset{k=1}{\overset{j}{\bigstar}}\Phi(x_k)\qquad\forall j:0\leq j\leq n.
\end{equation}
\item Each diagram remains invariant under the twist. In other words, a twist that acts simultaneously with one leg on all external fields becomes trivial. If we denote by $z$ another point not belonging to the diagram under study, this can be written as
\begin{equation}
    I_n\times\mathcal{F}_{xz}\left(\underset{k=1}{\overset{n}{\bigstar}}\Phi(x_k)\right)=I_n\times\left(\Delta^{(n)}\otimes1\right)(\mathcal{F})_{x_1\dots x_n z}\left( \underset{k=1}{\overset{n}{\bigstar}}\Phi(x_k) \right)=I_n\times\underset{k=1}{\overset{n}{\bigstar}}\Phi(x_k).
\end{equation}

\end{enumerate}

\paragraph{Proving the theorem} The theorem can be proven recursively. Starting from a tree-level diagram containing only one vertex, all tree-level  diagrams can be generated by recursively adding new vertices by simply contracting one external field of the existing diagram with one external field of the new interaction term. In order to obtain all planar loop diagrams, starting from any planar diagram (including all tree-level diagrams), we can contract two neighbouring external fields of a given planar diagram. This procedure generates all possible planar diagrams of a given theory in a systematic recursive way starting from the elementary interaction terms appearing in the action. In \cite{Meier:2023lku}, this procedure was used to prove the planar equivalence theorem in the context of deformations within the Poincar\'e algebra. Nevertheless, the proof purely relies on elementary properties of the propagators and interaction terms in the action of the deformed theory, which we will state explicitly below and argue that apply in our setting. For a proper presentation of the full recursive proof, we refer to \cite{Meier:2023lku}.

\paragraph{The propagators} For the recursive proof of the planar equivalence theorem to apply as presented in \cite{Meier:2023lku}, the propagator of the deformed theory has to satisfy the following conditions.
\begin{enumerate}
    \item The propagators of the deformed theory need to stay undeformed.
    \item Acting with one leg of the twist on one end of the propagator has to be related to the twist acting on the other end instead as
    \begin{equation}
    \bar f_x^\alpha(\Delta_\Phi(x-y))=S(\bar f_y^\alpha)(\Delta_\Phi(x-y)).
    \end{equation}
    This property is essential to move internal twists in a Feynman diagram. In fact it will allow to move a twist from one end of the propagator to the other, which is crucial for the recursive proof.
\end{enumerate}

\paragraph{The interactions} Similar to the discussion of the propagators, the deformed interaction terms have to satisfy the following conditions for the proof of the planar equivalence theorem to hold.
\begin{enumerate}
    \item Each interaction term has to be cyclic with respect to the star product. This for example ensures that the simplest diagrams, i.e. diagrams with only one vertex, satisfy the planar equivalence theorem.
    \item Similarly to the case of the propagator, acting with the twist on one external field is related to acting on all other external fields instead as
    \begin{equation}
        I_n\times\bar f^\alpha(\Phi(x_1))\star\underset{k=2}{\overset{n}{\bigstar}}\Phi(x_k)=I_n \times\Phi(x_1)\star S(\bar f_\alpha)\left(\underset{k=2}{\overset{n}{\bigstar}}\Phi(x_k)\right).
    \end{equation}
    This is necessary to move internal twists around, and in particular to finally move them to external fields instead and cancel remaining internal twists.
\end{enumerate}

Note that these properties do not directly apply to all kinds of propagators and interaction terms in the deformed theories. Taking a twist in the Poincar\'e algebra, a propagator of a component fermionic or vector field is not Lorentz invariant. Similarly, the interaction terms in component fields are only cyclic up to appropriate powers of $\mathcal R$-matrices in several representations. To overcome this problem in the Poincar\'e case, in \cite{Meier:2023lku} the Feynman diagrams were treated in an index-free formulation directly following the index-free formulation of the action. For example, the propagator of a gauge one-form is given as the Feynman propagator of the vector field appropriately multiplied by the basis one-forms coming with both vector fields:
\begin{equation}
    \Braket{A(x)A(y)}=\Braket{A_\mu(x)A_\nu(y)}\dd x^\mu\dd y^\nu=\frac{\dd x^\mu\dd y^\nu\eta_{\mu\nu}}{(x-y)^2}.
\end{equation}
The advantage of this notation is that in this way, all propagators are naturally Poincar\'e invariant, and hence condition 2 for the propagators directly apply for deformations of the Poincar\'e algebra. Similarly, in index-free form the vertices are manifestly Poincar\'e invariant and (graded) cyclic. Hence, they satisfy the preliminary conditions for the interaction terms as well.

In the following, we will adopt this formalism to twists including the dilatation generator. In particular, we will consider Wick contractions of the scaleless fields and the interaction terms written in terms of these new scaleless fields as well. While we will present explicitly the derivation for the scalar fields, the discussion can be easily extended to cover vector and fermionic fields in their scaleless and index-free form as well.

\paragraph{Proving the preliminary conditions}
In general, in the constructions above, the kinetic actions remain undeformed because of the cyclic star product and the undeformed Laplace operator. Hence, the propagators of massless theories will remain undeformed as well. Note that the mass terms for fundamental matter fields introduced above are truly deformed. Hence, for deformations involving $\mathcal D$, massive propagators appear to be deformed and the planar equivalence theorem does not apply for massive theories, which is why we will only consider massless theories in the remainder of this section.

To prove the second condition for propagators in the example of scalar fields, we will first construct the scaleless propagators. These are defined as corresponding to the Wick contraction of the scaleless fields $\varphi$.
\begin{equation}
    \Delta_\varphi(x-y)=\Braket{\varphi(x)\varphi(y)}=H_x^{-1}H^{-1}_y\Braket{\phi(x)\phi(y)}=\frac{H_x^{-1}H^{-1}_y}{(x-y)^2},
\end{equation}
where we add an index to $H$ to keep track to which field it corresponds to. This propagator is clearly invariant under Poincar\'e transformations $X$ as
\begin{equation}
    (X_x+X_y)\Delta_\varphi(x-y)=0.
\end{equation}
It is furthermore invariant under the action of the dilatation $\mathcal{D}$ as
\begin{equation}
\begin{aligned}
    (\mathcal{D}_x+\mathcal{D}_y)\Delta_\varphi(x-y)&=(-ix^\mu\partial^x_\mu-iy^\mu\partial_\mu^y+iH_x\partial_{H_x}+iH_y\partial_{H_y})\frac{H_x^{-1}H^{-1}_y}{(x-y)^2}\\
    &=0.
\end{aligned}
\end{equation}
We can lift this equation recursively to any element $X$ of the full universal enveloping algebra of rescalings and Poincar\'e symmetries as
\begin{equation}
     (X_x-S(X)_y)\Delta_\varphi(x-y),
\end{equation}
where $S$ again denotes the antipode of the Hopf algebra.
As $f^\alpha$, appearing in the definition of the twist, is in the universal enveloping algebra, the second condition on the propagators is satisfied.

We will finally address the condition for the proof of the planar equivalence theorem on the interaction terms. As already discussed, the twist version of the unimodularity condition in \eqref{eq:intCyc} ensures cyclicity of the star product under integration, if the undeformed integral is invariant under scale and Poincar\'e transformations. This is the case for all interactions that are conformally invariant in the undeformed theory. Similarly, partial integration ensures the second property for scale-invariant interactions.

By constructing the scale invariant propagators for fermionic and vector fields as
\begin{equation}
\begin{aligned}
    \Braket{A(x)A(y)}&=\frac{\eta_{\mu\nu}\dd x^\mu\dd x^\nu}{(x-y)^2}\\
    \Braket{\Psi(x)\bar\Psi(y)}&=\partial_{\alpha\dot\alpha}\frac{s_x^\alpha \bar s_y^{\dot\alpha}H_x^{-\frac{3}{2}}H_y^{-\frac{3}{2}}}{(x-y)^2},
\end{aligned}
\end{equation}
the preliminary conditions of the proof of the planar equivalence theorem apply to all possible massless theories constructed via the approach in section \ref{sec:gauge-theories}. Hence, the planar equivalence theorem for Feynman diagrams also applies to deformations of conformal theories by twists including the dilatation generator.
\vspace{12pt}

\section{Conclusions}
We constructed twisted non-commutative versions of star-gauge theories including adjoint and fundamental matter fields. The twists are generated by Drinfel'd twists built from the dilatation generator and the Poincar\'e algebra. Importantly, to respect gauge invariance, we restricted ourselves to twists with a unimodular classical r-matrix. Our construction was carried out following two approaches. 

The first approach generalises the construction given in \cite{Meier:2023lku} to cases beyond the Poincar\'e algebra, now including scale transformations. We did so by introducing an auxiliary object $H$ that carries a scale. After generalizing the Hodge duality using $H$ and rewriting the undeformed action in terms of scale- and index-free objects, the non-commutative deformation was obtained by replacing products by star products. The theory is then invariant under star-gauge symmetries, as the deformed generalised Hodge operator is star-linear and the unimodularity condition for the r-matrix leads to a cyclic star product under integration. Furthermore, we were able to include massless fundamental and adjoint matter fields as well as massive fundamental matter fields. This is enough to define a star-gauge invariant non-commutative version of $\mathcal{N}=4$ SYM for all twists that have a unimodular r-matrix and are built from the scale and Poincar\'e algebra. We proved the applicability of the planar equivalence theorem for Feynman diagrams \cite{FILK199653,Meier:2023lku} for this large class of deformations, which allows one to map the deformed planar Feynman diagrams to their undeformed counterparts up to star products acting on the external positions only. Furthermore, we proved that the deformed theory is  invariant under twisted scale- and Poincar\'e transformations, as may be expected from potential dual candidates of the Yang-Baxter deformed $AdS_5\times S^5$ string. 

At the same time, we  introduced also a second approach to twisted field theories. Instead of twisting the algebra of vector fields acting on the space of fields, we changed the perspective to directly twist the symmetries of a given theory. In this approach, the twist acts as active transformations on the fields. This leaves the differential calculus intact and leads to an alternative formulation of non-commutative gauge theories. We also showed that the resulting action is equivalent to the corresponding action constructed  via the Lie-derivative picture of the first approach.

The main part of this work focuses on the construction of classical actions. There are therefore various open questions on the quantum behaviour of the theory, for example under renormalization, such as UV/IR mixing both for the supersymmetric and the general case. UV/IR mixing is a common feature for non-commutative field theories, as discussed  for the Groenewold-Moyal case in \cite{Szabo:2001kg}, for certain theories on $\kappa$-Minkowski space in \cite{Grosse:2005iz} and on $\lambda$-Minkowski space in \cite{DimitrijevicCiric:2018blz}, for instance. In all these cases, the deformation parameter itself is dimensionful, and one may expect that this feature may play a role in the presence of the UV/IR mixing. Importantly, in the case of quadratic twists, such as for example $e^{i\lambda D\wedge M_{23}}$, the deformation parameter is dimensionless instead, and it is worth asking whether UV/IR mixing will appear or not. Moreover, it is also known that supersymmetric theories do not admit UV/IR mixing even in the Groenewold-Moyal case \cite{Szabo:2001kg}. Other natural open questions include the construction of  the Seiberg-Witten map for the resulting theories, and  if the twisted symmetries, especially scale transformations, will in general survive at arbitrary loop level. Importantly, given that scaling dimensions of operators can have anomalous dimensions and that operators can mix under the action of the dilatation operator, it is natural to wonder whether in the quantum setup the twist should be defined still using the classical scaling dimensions of the fields, or if quantum corrections to the action of the dilatation operator should be included.

Having in mind our motivation of studying integrable deformations of the AdS/CFT correspondence, the constructed deformations of $\mathcal{N}=4$ SYM are conjectured to be dual to homogeneous Yang-Baxter deformations of the $AdS_5\times S^5$ string \cite{vanTongeren:2015uha,vanTongeren:2016eeb}. In this setup, we are interested in the deformation of the underlying integrable structures, particularly in the twisting of the spin-chain picture appearing in the planar two-point functions of gauge invariant operators. This twist has been observed in the context of marginal deformations of $\mathcal{N}=4$ SYM in \cite{Fokken:2013mza,Beisert:2005if} and has been found for dipole-like deformations more recently in \cite{Guica:2017mtd,Meier:2025tjq}. A natural next candidate to study would be the scale-dipole deformation, where the twist is abelian and is built from an internal R-charge and the dilatation generator. This case lacks an invariant plane, unlike in the angular dipole deformation \cite{Meier:2025tjq}, but it is still built from Cartan charges. It can therefore be a perfect bridge from the study of dipole deformations towards proper non-commutative theories, such as the Lorentz deformation \cite{Meier:2023kzt}. In the context of identifying spin-chain integrability, the planar equivalence theorem has been an important tool \cite{Meier:2025tjq,Guica:2017mtd}.

One of our aims is to  extend the applicability of the construction of the twist non-commutative deformations to the full conformal, and even to the full superconformal, algebra. In particular, we will present the construction with the approach of the active transformations   in a work to appear\cite{next-paper}, which will allow us to construct all conjectured dual candidates to homogeneous Yang-Baxter deformations.  Such deformed gauge theories would then also include the duals to unimodular extensions of the Jordanian deformations \cite{vanTongeren:2019dlq,Borsato:2022ubq}. These were studied from the point of view of the string sigma-model in \cite{Borsato:2022drc,Borsato:2024sru}, and  integrable spin-chain Hamiltonians describing a sector of the gauge theories were also recently proposed and studied in~\cite{Borsato:2025smn,deLeeuw:2025sfs,Driezen:2025dww,Driezen:2025izd}. 

Besides integrability, it would be interesting to investigate the connection between the graded cyclicity of the star product under integration and the Weyl invariance of the Yang-Baxter deformed string theory, as both are ensured by the unimodularity condition on the classical r-matrix. In this setup, it would be interesting to see explicitly how adding fermionic generators to promote a twist to a unimodular one can  ensure cyclicity of the star product under integration.

The non-commutative versions of $\mathcal{N}=4$ SYM are expected to be classically integrable in the sense of having a Yangian symmetry along the lines of \cite{Beisert:2017pnr,Beisert:2018zxs,Garus:2018ajc}. In the case of the deformed theories, the Yangian is expected to act via the twisted coproduct instead of the untwisted structure. This is similar to the case of the $\beta$-deformation\cite{Beisert:2024wqq} and the (fishnet limit of the) $\gamma_i$ deformations \cite{Beisert:2025mtn}.

In the marginal $\gamma_i$ deformations, which are generated by a twist built from all the Cartan charges of the internal R-symmetry, there exists a strong deformation limit, also known as the fishnet limit \cite{Gurdogan:2015csr}, which results in an integrable, non-hermitian matrix-valued scalar field theory. In terms of integrability, this theory has been a prime example for studying aspects of integrability in field theories. It is an old hope to generalise this strong deformation limit to the deformation of $\mathcal{N}=4$ SYM that twists the theory by all Cartan charges of the superconformal algebra, but this idea has been so far inaccessible.  By including scale transformations in the twist, our work permits the construction of the desired twist of  $\mathcal{N}=4$ SYM. This opens up the possibility to study all possible strong deformation limits, and in principle non-commutative cousins of the fishnet theory which are expected to be part of a larger family of non-hermitian integrable models. 

Beyond Yang-Mills theory and the $AdS_5$/$CFT_4$ correspondence, it would be interesting to investigate this kind of non-commutativities also in the context of Chern-Simons-like gauge theories, especially as they naturally appear in the context of ABJM theory in the $AdS_4$/$CFT_3$ correspondence \cite{Aharony:2008ug, Hosomichi:2008jb}.

\section*{Acknowledgements}
We thank Stijn van Tongeren for precious comments on the draft.
This work was supported by the grants RYC2021-032371-I (funded by the European Union ``NextGenerationEU''/PRTR and by MCIN/AEI/10.13039/501100011033), 2023-PG083 (with reference code ED431F 2023/19 funded by Xunta de Galicia),  PID2023-152148NB-I00 (funded by AEI-Spain), the Mar\'ia de Maeztu grant CEX2023-001318-M (funded by MICIU/AEI /10.13039/501100011033), the CIGUS Network of Research Centres, and the European Union.

\appendix
\section{Conventions}\label{app:conv}
Here we collect our conventions on the generators of the Poincar\'e algebra extended by the scale transformation. When representing them as Lie derivatives $\mathcal L_X$ with $X=\xi^\mu\partial_\mu$, we will take
\begin{equation}
    p_\mu=-i\partial_\mu,\qquad M_{\mu\nu}=2i\, x_{[\mu}\partial_{\nu]}=i(x_\mu\partial_\nu-x_\nu\partial_\mu),\qquad D=-i\, x^\mu\partial_\mu,
\end{equation}
respectively for translations, Lorentz rotations and the scale transformation,
so that their commutation relations are\footnote{Notice that these relations imply $ e^{i\epsilon\, D}p_\mu e^{-i\epsilon\, D}=e^{-\epsilon}p_\mu$ and $e^{\frac{i}{2}\omega^{\alpha\beta}M_{\alpha\beta}}p_\mu e^{-\frac{i}{2}\omega^{\alpha\beta}M_{\alpha\beta}} = \Lambda_\mu{}^\nu p_\nu$, where $\Lambda_\mu{}^\nu$ is the Lorentz matrix defined in~\eqref{eq:Lor-mat}.}
\begin{equation}\begin{aligned}
     &[M_{\mu\nu},M_{\rho\sigma}]=-i(\eta_{\mu\rho}M_{\nu\sigma}-\eta_{\nu\rho}M_{\mu\sigma}+\eta_{\nu\sigma}M_{\mu\rho}-\eta_{\mu\sigma}M_{\nu\rho}),\\
     &[M_{\mu\nu},p_\rho]=-i(\eta_{\rho\mu}p_\nu-\eta_{\rho\nu}p_\mu),\qquad [D,p_\mu]=i\, p_\mu.
\end{aligned}
\end{equation}
These symmetry transformations will have a \emph{finite} action given respectively by
\begin{equation}
    e^{ia^\mu p_\mu},\qquad e^{\frac{i}{2}\omega^{\mu\nu}M_{\mu\nu}},\qquad e^{i\epsilon D},
\end{equation}
where $a^\mu$ are the parameters for the translations, $\omega^{\mu\nu}$ those for Lorentz rotations and $\epsilon$ for the scale transformation. The conventions are chosen such that under the finite transformations the coordinates change as
\begin{equation}
    x'{}^\mu = x^\mu+a^\mu,\qquad
    x'{}^\mu = \Lambda^\mu{}_\nu x^\nu,\qquad
    x'{}^\mu = e^\epsilon x^\mu,
\end{equation}
respectively, where we defined the Lorentz matrix
\begin{equation}\label{eq:Lor-mat}
    \Lambda^\mu{}_\nu=(e^{\omega})^\mu{}_\nu.
\end{equation}
The finite actions of the symmetry transformations on functions and generic fields are then
\begin{equation}
    e^{ia^\mu p_\mu}\Phi(x)=\Phi(x+a),\qquad e^{\frac{i}{2}\omega^{\mu\nu}M_{\mu\nu}}\Phi(x) =\Phi(\Lambda x),\quad ,\qquad e^{i\epsilon D}\Phi(x) = \Phi(e^\epsilon x).
\end{equation}
It may be useful to adopt different notions of symmetry transformations on fields. First, one may want to work with \emph{passive} transformations where the field is unchanged while the coordinates do change. At the infinitesimal level, then, if $x'\approx x+\xi$, passive transformations are captured by the above Lie derivatives
\begin{equation}
    \Phi(x')-\Phi(x)\approx \xi^\mu\partial_\mu\Phi(x).
\end{equation}
Alternatively, one may want to work with \emph{active} transformations $ \Phi'(x)-\Phi(x)$ where the fields transform while the coordinates remain unchanged. Transformation rules for the fields are fixed by requiring that the symmetry leave the action invariant. Under translations, all fields are assumed to transform as $\Phi'(x')=\Phi(x)$, so that  the active transformations are $ \Phi'(x)-\Phi(x)=\Phi(x-a)-\Phi(x)\approx -a^\mu\partial_\mu\Phi=-\mathcal L_{ia^\mu p_\mu}\Phi$. If we call $\hat p_\mu$ the representation as active transformation of translations, we may then identify $\hat p_\mu=-p_\mu$. This change of sign relating the active transformations with the Lie-derivative representation appearing in the passive transformation generalises to the other generators in the following way. 

Under a scale transformation with $x'=\lambda x$, where we defined $\lambda=e^\epsilon$, a generic field transforms as 
    \begin{equation}
        \Phi'(x')=\lambda^{-\Delta_\Phi}\Phi(x),
    \end{equation}
    where $\Delta_\Phi$ is the scaling dimension. 
    The active transformation is then $\Phi'(x)-\Phi(x)=\lambda^{-\Delta_\Phi}\Phi(\lambda^{-1}x)-\Phi(x)\approx -\epsilon x^\mu\partial_\mu\Phi - \epsilon\Delta_\Phi\Phi$, and if we identify this result with $i\epsilon\hat D\Phi$ we can define the representation of the scale symmetry as an active transformation by
    \begin{equation}
        \hat D = i\, x^\mu\partial_\mu + i\Delta_\Phi\Phi\frac{\delta}{\delta\Phi}=-D + i\Delta_\Phi\Phi\frac{\delta}{\delta\Phi}.
    \end{equation}
    The finite active transformation then is the desired one, $e^{i\epsilon\hat D}\Phi(x) = e^{-\epsilon\Delta_\Phi}\Phi(e^{-\epsilon}x)$.
    For a scalar field $\phi$, a Dirac/Weyl fermion $\psi$ and a vector $A_\mu$ the scaling dimensions are
    \begin{equation}
        \Delta_\phi=1,\qquad \Delta_\psi=3/2,\qquad \Delta_A=1,
    \end{equation}
    so that we have
    \begin{equation}
        \begin{aligned}
            \hat D(\phi)&=ix^\mu\partial_\mu\phi+i\phi,\\
            \hat D(\psi)&=ix^\mu\partial_\mu\psi+i\tfrac{3}{2}\psi,\\
            \hat D(A_\nu)&=ix^\mu\partial_\mu A_\nu+iA_\nu.
        \end{aligned}
    \end{equation}
Similarly, we may consider the Lorentz transformation $x'{}^\mu=\Lambda^\mu{}_\nu x^\nu$ and define $\hat M_{\mu\nu}$ generators that implement the active transformations on scalars, Dirac/Weyl fermion and vectors as
    \begin{equation}
        \begin{aligned}
            \hat M_{\mu\nu}(\phi)&=-2i\, x_{[\mu}\partial_{\nu]}\phi,\\
            \hat M_{\mu\nu}(\psi)&=-2i\, x_{[\mu}\partial_{\nu]}\psi+\Sigma_{\mu\nu}\psi,\\
            \hat M_{\mu\nu}(A_\rho)&=-2i\, x_{[\mu}\partial_{\nu]}A_\rho -2i\eta_{\rho[\mu}A_{\nu]},
        \end{aligned}
    \end{equation}
    where $\Sigma_{\mu\nu}=-\tfrac{i}{4}[\gamma_\mu,\gamma_\nu]$ and $\gamma_\mu$ are gamma-matrices. In the Weyl representation one can write
    \begin{equation}
    \gamma^\mu = \left(\begin{array}{cc}
      \mathbf 0   & \sigma^\mu \\
         \bar \sigma^\mu&\mathbf  0 
    \end{array}\right),
\end{equation}
with 
\begin{equation}
    \sigma^\mu = (\mathbf 1,\sigma^i),\qquad\qquad
    \bar\sigma^\mu = (\mathbf 1,-\sigma^i),
\end{equation}
so that 
\begin{equation}
    \begin{aligned}
        \Sigma^{\mu\nu} = -i\left(\begin{array}{cc}
            \sigma^{\mu\nu} & \mathbf 0 \\
            \mathbf 0 & \bar\sigma^{\mu\nu}
        \end{array}\right) ,
    \end{aligned}
\end{equation}
with 
\begin{equation}
    \sigma^{\mu\nu}=\frac{1}{4}(\sigma^\mu\bar\sigma^\nu-\sigma^\nu\bar\sigma^\mu),\qquad\qquad
    \bar\sigma^{\mu\nu}=\frac{1}{4}(\bar\sigma^\mu\sigma^\nu-\bar\sigma^\nu\sigma^\mu).
\end{equation}
Then the active transformations on Weyl fermions with positive/negative chirality are
\begin{equation}
    \hat M_{\mu\nu}(\psi_+)=-2i\, x_{[\mu}\partial_{\nu]}\psi_+-i\sigma_{\mu\nu}\psi_+,\qquad
    \hat M_{\mu\nu}(\psi_-)=-2i\, x_{[\mu}\partial_{\nu]}\psi_--i\bar\sigma_{\mu\nu}\psi_-.
\end{equation}
To conclude this section, let us point out that the representation as active transformations of the symmetry generators amounts to an automorphism $X\to -X$ of the Lie algebra, so that the commutation relations are
\begin{equation}\begin{aligned}
     &[\hat M_{\mu\nu},\hat M_{\rho\sigma}]=i(\eta_{\mu\rho}\hat M_{\nu\sigma}-\eta_{\nu\rho}\hat M_{\mu\sigma}+\eta_{\nu\sigma}\hat M_{\mu\rho}-\eta_{\mu\sigma}\hat M_{\nu\rho}),\\
     &[\hat M_{\mu\nu},\hat p_\rho]=i(\eta_{\rho\mu}\hat p_\nu-\eta_{\rho\nu}\hat p_\mu),\qquad [\hat D,\hat p_\mu]=-i\, \hat p_\mu.
\end{aligned}
\end{equation}
\section{Deformed Levi-Civita symbol for non scale invariant basis forms}\label{app:alternative-epsilon}
In section \ref{sec:twist-diff-calc}, we defined Hodge duality naturally on scale-invariant basis forms via a deformed Levi-Civita symbol using the scale-invariant basis forms $\dd\xi^\mu$ only. We can change the perspective and can derive a Hodge duality and deformed Levi-Civita symbol for the standard basis forms $\dd x^\mu$. We start by defining the deformed Levi-Civita symbol in the standard basis as
\begin{equation}
    \varepsilon_x^{\mu\nu\rho\sigma}\dd^4x=\dd x^\mu \wedge_\star \dd x^\nu \wedge_\star \dd x^\rho \wedge_\star \dd x^\sigma.
\end{equation}
In order to contrast the Levi-Civita symbol $\varepsilon_x^{\mu\nu\rho\sigma}$ to its counterpart $\varepsilon_\star^{\mu\nu\rho\sigma}$, we will repeat the definition of the latter as
\begin{equation}
\label{eq:defEpsilonInApp}
    \varepsilon_\star^{\mu\nu\rho\sigma}\dd^4\xi=\dd \xi^\mu\wedge_\star\dd \xi^\nu\wedge_\star\dd \xi^\rho\wedge_\star\dd \xi^\sigma,
\end{equation}
see also equation \eqref{eq:defEpsilon}. We start expressing the scaleless volume form $\dd^4\xi$ in terms of the standard basis:
\begin{equation}
    \dd^4\xi=\dd \xi^0 \wedge \dd\xi^1 \wedge \dd\xi^2 \wedge \dd\xi^3=\left(H\star\dd x^0\right)\wedge\left(H\star\dd x^1\right)\wedge\left(H\star\dd x^2\right)\wedge\left(H\star\dd x^3\right).
\end{equation}
The star products between $H$ and the basis forms $\dd x^\mu$ can be expressed as normal products up to the twist $\tensor{\left.\bar {\tilde F}_{op}\right.}{_\nu^\mu}$ and hence
\begin{equation}
\begin{aligned}
    \dd^4\xi&=\tensor{\left.\bar {\tilde F}_{op}\right.}{_\mu^0}\tensor{\left.\bar {\tilde F}_{op}\right.}{_\nu^1}\tensor{\left.\bar {\tilde F}_{op}\right.}{_\rho^2}\tensor{\left.\bar {\tilde F}_{op}\right.}{_\sigma^3}H^4\dd x^\mu\wedge \dd x^\nu\wedge \dd x^\rho\wedge \dd x^\sigma\\
    &=\tensor{\left.\bar {\tilde F}_{op}\right.}{_\mu^0}\tensor{\left.\bar {\tilde F}_{op}\right.}{_\nu^1}\tensor{\left.\bar {\tilde F}_{op}\right.}{_\rho^2}\tensor{\left.\bar {\tilde F}_{op}\right.}{_\sigma^3}\varepsilon^{\mu\nu\rho\sigma}H^4\dd^4 x.
\end{aligned}
\end{equation}
Note, that $\tensor{\left.\bar {\tilde F}_{op}\right.}{_\nu^\mu}$ is a Poincar\'e transformation matrix and hence
\begin{equation}
    \tensor{\left.\bar {\tilde F}_{op}\right.}{_\mu^\alpha}\tensor{\left.\bar {\tilde F}_{op}\right.}{_\nu^\beta}\tensor{\left.\bar {\tilde F}_{op}\right.}{_\rho^\gamma}\tensor{\left.\bar {\tilde F}_{op}\right.}{_\sigma^\delta}\varepsilon^{\mu\nu\rho\sigma}=\varepsilon^{\alpha\beta\gamma\delta},
\end{equation}
which leads to
\begin{equation}
    \dd^4\xi=H^4\dd x^0\wedge\dd x^1\wedge\dd x^2\wedge\dd x^3=H^4\star\dd^4 x.
\end{equation}
The right-hand-side of equation \eqref{eq:defEpsilonInApp} results in
\begin{equation}
    \dd \xi^\mu \wedge_\star \dd \xi^\nu \wedge_\star \dd \xi^\rho \wedge_\star \dd \xi^\sigma= \tensor{\tilde R}{_{\rho'}^\rho}\tensor{\tilde R}{_{\nu'}^\nu}\tensor{\tilde R}{_{\nu''}^{\nu'}}\tensor{\tilde R}{_{\mu'}^\mu}\tensor{\tilde R}{_{\mu''}^{\mu'}}\tensor{\tilde R}{_{\mu'''}^{\mu''}}H^4 \star \dd x^{\mu'''}\wedge_\star\dd x^{\nu''}\wedge_\star\dd x^{\rho'}\wedge_\star\dd x^{\sigma}
\end{equation}
and hence
\begin{equation}
    \varepsilon_\star^{\mu\nu\rho\sigma}H^4\star\dd^4x=\tensor{\tilde R}{_{\rho'}^\rho}\tensor{\tilde R}{_{\nu'}^\nu}\tensor{\tilde R}{_{\nu''}^{\nu'}}\tensor{\tilde R}{_{\mu'}^\mu}\tensor{\tilde R}{_{\mu''}^{\mu'}}\tensor{\tilde R}{_{\mu'''}^{\mu''}}\varepsilon_x^{\mu'''\nu''\rho'\sigma}H^4\star\dd^4 x,
\end{equation}
which determines how to change the basis in the deformed Levi-Civita symbol.

From the commutation relations of basis forms, the deformed Levi-Civita symbol is $\mathcal{R}$-antisymmetric in all its indices:
\begin{equation}
\varepsilon_x^{\mu\nu\rho\sigma}=-\tensor{R}{_\alpha^\mu_\beta^\nu}\varepsilon_x^{\beta\alpha\rho\sigma}=-\tensor{R}{_\alpha^\nu_\beta^\rho}\varepsilon_x^{\mu\beta\alpha\sigma}=-\tensor{R}{_\alpha^\rho_\beta^\sigma}\varepsilon_x^{\mu\nu\beta\alpha}.
\end{equation}
Moreover, conjugation of the star-wedge product results in
\begin{equation}
\varepsilon_x^{\mu\nu\rho\sigma}=\overline{\varepsilon_x^{\sigma\rho\nu\mu}}.
\end{equation}
Note that, as the volume form is charged under scale transformations, the deformed Levi-Civita symbol is not graded-cyclic anymore.  Reducing the setup to twists in the Poincar\'e algebra, however, one recovers cyclicity. \\

The volume form $ \dd^4 x $ is invariant under Poincaré transformations but possesses an inherent scale. As a result, functions do not star-commute with the volume form. To account for this non-commutativity, we introduce an additional representation of the $\mathcal{R}$-matrix. Since this version of the $\mathcal{R}$-matrix solely encodes the scaling behavior of the basis forms, it carries no indices. Consequently, the commutation relation between functions and the volume form is given by  
\begin{equation}  
\dd^4 x \star f = \tilde{R}^4 f \star \dd^4 x.  
\end{equation}
By star multiplying \eqref{eq:defEpsilon} from the right with a function $f$ and then using the commutation relation of functions with the volume form on the left-hand side and of functions and basis forms on the right-hand side, we can derive the identity
\begin{equation}
\label{eq:epCyc}
\varepsilon_x^{\mu\nu\rho\sigma}\tilde{R}^4=\varepsilon_x^{\alpha\beta\gamma\delta}\tensor{R}{_\alpha^\mu}\tensor{R}{_\beta^\nu}\tensor{R}{_\gamma^\rho}\tensor{R}{_\delta^\sigma}.
\end{equation}
This equation holds as an operator equation. We can evaluate it explicitely on a basis one-form, which results in
\begin{equation}
\label{eq:epCyc2}
    \varepsilon_x^{\mu\nu\rho\sigma}\tensor{\left.\tilde{R}^4\right.}{_\kappa^\tau}=\varepsilon_x^{\alpha\beta\gamma\delta}\tensor{R}{_\alpha^\mu_\kappa^{\kappa'}}\tensor{R}{_\beta^\nu_{\kappa'}^{\tau'}}\tensor{R}{_\gamma^\rho_{\tau'}^{\tau^{''}}}\tensor{R}{_\delta^\sigma_{\tau^{''}}^\tau}.
\end{equation}
The latter identity is particularly useful for obtaining a modified cyclicity identity  for the deformed Levi-Civita symbol. Let us start from equation \eqref{eq:defEpsilon} and use the $\mathcal{R}$-antisymmetry of the star-wedge product:
\begin{equation}
\begin{aligned}
\varepsilon_x^{\mu\nu\rho\sigma}\dd^4x&=-\tensor{R}{_{\mu'}^\mu_{\tau'}^\tau}\tensor{R}{_{\nu'}^\nu_\tau^{\sigma'}}\tensor{R}{_{\rho'}^\rho_{\sigma'}^\sigma}\dd x^{\tau'} \wedge_\star \dd x^{\mu'} \wedge_\star \dd x^{\nu'} \wedge_\star \dd x^{\rho'}\\
&=-\tensor{R}{_{\kappa}^\alpha_{\alpha}^{\tau'}}\tensor{R}{_{\mu'}^\mu_{\tau'}^\tau}\tensor{R}{_{\nu'}^\nu_\tau^{\sigma'}}\tensor{R}{_{\rho'}^\rho_{\sigma'}^\sigma}\dd x^{\kappa} \wedge_\star \dd x^{\mu'} \wedge_\star \dd x^{\nu'} \wedge_\star \dd x^{\rho'}\\
&=-\tensor{R}{_{\kappa}^\alpha_{\alpha}^{\tau'}}\tensor{R}{_{\mu'}^\mu_{\tau'}^\tau}\tensor{R}{_{\nu'}^\nu_\tau^{\sigma'}}\tensor{R}{_{\rho'}^\rho_{\sigma'}^\sigma}\varepsilon^{\kappa\mu'\nu'\rho'}_\star\dd^4 x\\
&=-\tensor{\left(\tilde{R}^4\right)}{_\alpha^\sigma}\varepsilon_x^{\alpha\mu\nu\rho} \dd^4x,
\end{aligned}
\end{equation}
where equation \eqref{eq:RMatContraction} was used in the second step and equation \eqref{eq:epCyc2} in the final step. This is in contrast to the $\varepsilon_\star$ in the main text, which is graded cyclic, cf. \eqref{eq:grad-cycl-eps-star}. In particular, this is the main reason for using the scaleless Levi-Civita $\varepsilon_\star$ along with the scaleless basis forms to define Hodge duality rather than $\varepsilon_x$ and the canonical basis one-forms $\dd x^\mu$ instead.
This formula should be compared with the property of graded cyclicity of $\epsilon_\star^{\mu\nu\rho\sigma}$, see~\eqref{eq:grad-cycl-eps-star}. Given that the modified cyclicity of $\varepsilon_x^{\mu\nu\rho\sigma}$ is more cumbersome to work with, in the main text we always use $\epsilon_\star^{\mu\nu\rho\sigma}$ as the definition of the deformed Levi-Civita tensor.

\section{Basis spinors}\label{app:BasisSpinors}
In this appendix, we will introduce our convention on the basis spinors needed for the Lie derivative based deformation of fermionic fields. Our conventions will be aligned with \cite{Meier:2023lku}. The basis left- and right-handed Weyl spinors $s^\alpha$ and $\bar s^{\dot\alpha}$ will capture the Lie-derivative notion of  a vector field acting on a spinor. In this sense, they behave like basis one-forms and are hence also referred to as ``half-forms''. A generic vector field $X$ acts on the basis spinors via the spinor lie derivative \cite{SpinorLie}:
\begin{equation}
    \begin{aligned}
         \mathcal{L}_Xs_\alpha&=-\frac{1}{8}(\partial_\mu X_\nu-\partial_\nu X_\mu)\sigma^{\mu}_{\alpha\dot\alpha}\sigma^{\nu\dot\alpha\beta}\psi_\beta\\
        \mathcal{L}_X\bar s^{\dot\alpha}&=-\frac{1}{8}(\partial_\mu X_\nu-\partial_\nu X_\mu)\sigma^{\mu\dot\alpha\alpha}\sigma^{\nu}_{\alpha\dot\beta}\bar s^{\dot\beta}.
    \end{aligned}
\end{equation}
For the Levi-Civita symbol, we choose
\begin{equation}
    \varepsilon^{12}=\varepsilon^{\dot 1\dot2}=\varepsilon_{21}=\varepsilon_{\dot2\dot1}=1,
\end{equation}
which can be used to lower and raise indices on spinors as follows
\begin{equation}
\begin{aligned}
    s_\alpha&=\varepsilon_{\alpha\beta}s^\beta & \qquad s^\alpha&=\varepsilon^{\alpha\beta}s_\beta\\
    \bar{s}_{\dot\alpha}&=\varepsilon_{\dot\alpha\dot\beta}\bar \bar s^{\dot\beta} & \qquad
    \bar s^{\dot\alpha}&=\varepsilon^{\dot\alpha\dot\beta}\bar s_{\dot\beta}
    \end{aligned}
\end{equation}
Hence, the product of two basis spinors is given by
\begin{equation}
\begin{aligned}
    s^\alpha s^\beta&=\frac{1}{2}\varepsilon^{\alpha\beta}s^\gamma s_\gamma
    &\qquad\bar{s}_{\dot\alpha}\bar{s}_{\dot\beta}&=\varepsilon_{\dot\alpha\dot\beta}\bar s_{\dot\gamma}\bar s^{\dot\gamma}.
\end{aligned}
\end{equation}
Furthermore, we define Grassmann integration as
\begin{equation}
    \int\dd^2s~s^\alpha s^\beta=\varepsilon^{\alpha\beta}\qquad\int\dd^2\bar s~\bar s^{\dot\alpha}\bar s^{\dot\beta}=-\varepsilon^{\dot\alpha\dot\beta}.
\end{equation}
Hence, integrating over the basis spinors is equivalent to the contraction between two spinors:
\begin{equation}
\begin{aligned}
    \int\dd^2s~\psi\chi&=\int\dd^2s~s_\beta s_\alpha \psi^\alpha\chi^\beta&=\psi^\alpha\chi_\alpha\\
    \int \dd^2\bar s~\bar \psi\bar\chi&=-\int\dd^2\bar s~\bar s^{\dot \alpha}\bar s^{\dot \beta}\bar\psi_{\dot\alpha}\bar\chi_{\dot\beta}&=\psi_{\dot\alpha}\bar\chi^{\dot\alpha}.
 \end{aligned}   
\end{equation}
In the deformed setup it will be useful to consider also star products between basis spinors. As the Lie derivative with respect to scale transformations is trivial on the basis spinors, the twist reduces to a pure twist in the Poincar\'e algebra when acting on basis spinors only. Furthermore, for all twists in the Poincar\'e algebra, that lead to a cyclic star product under integration, the star product between two left- (right-)handed basis spinors is trivial respectively. Nonetheless, the star product between a left and a right handed basis spinor is not trivial. We can express the twist acting on basis spinors with a set of spinor indices for each leg. Hence,
\begin{equation}
    \begin{aligned}
        f\star s^\alpha&=\left(\tensor{\left. \bar F_{op} \right.}{_\beta^\alpha}f\right)s^\beta & f\star\bar s^{\dot\alpha}&=\left(\tensor{\left. \bar F_{op} \right.}{_{\dot\beta}^{\dot\alpha}}f\right)s^{\dot\beta}\\
        s^\alpha\star\bar s^{\dot\alpha}&=\tensor{\bar F}{_\beta^\alpha_{\dot\beta}^{\dot\alpha}}s^\beta\bar s^{\dot\beta}& s^\alpha\star\bar s^{\dot\alpha}&=\tensor{\bar F}{_{\dot\beta}^{\dot\alpha}_\beta^\alpha}s^\beta\bar s^{\dot\beta}.
    \end{aligned}
\end{equation}
Similarly, basis spinors star commute up to $\mathcal R$-matrix contributions like
\begin{equation}
\begin{aligned}
    f\star s^\alpha&=s^\beta\star\left(\tensor{R}{_\beta^\alpha}f\right) & f\star\bar s^{\dot\alpha}&=s^{\dot\beta}\star\left(\tensor{R}{_{\dot\beta}^{\dot\alpha}}f\right)\\
        s^\alpha\star\bar s^{\dot\alpha}&=\tensor{R}{_\beta^\alpha_{\dot\beta}^{\dot\alpha}} \bar s^{\dot\beta}\star s^\beta& s^\alpha\star\bar s^{\dot\alpha}&=\tensor{R}{_{\dot\beta}^{\dot\alpha}_\beta^\alpha}\bar s^{\dot\beta} \star s^\beta
\end{aligned}
\end{equation}
Furthermore, basis spinors obey the commutation relations with $H$ as
\begin{equation}
    \begin{aligned}
        H\star s^\alpha&=\tensor{\tilde R}{_\beta ^\alpha}s^\beta \star H\\
        H\star \bar s^{\dot\alpha}&=\tensor{\tilde R}{_{\dot\beta} ^{\dot\alpha}}\bar s^{\dot\beta} \star H.
    \end{aligned}
\end{equation}



\begin{thebibliography}{10}
\ifx\href\asklfhas\newcommand{\href}[2]{#2}\fi
\ifx\arxivref\asklfhas\newcommand{\arxivref}[2]{\href{http://arxiv.org/abs/#1}{#2}}\fi
\ifx\doiref\asklfhas\newcommand{\doiref}[2]{\href{http://dx.doi.org/#1}{#2}}\fi
\raggedright
\small
\parskip 0pt

\bibitem{Snyder:1946qz}
H.~S.~Snyder,
\textit{``{Quantized space-time}''},
\textsf{\doiref{10.1103/PhysRev.71.38}{Phys.~Rev.~71,~38~(1947)}}.

\bibitem{Doplicher:1994tu}
S.~Doplicher, K.~Fredenhagen and J.~E.~Roberts,
\textit{``{The Quantum structure of space-time at the Planck scale and quantum
  fields}''},
\textsf{\doiref{10.1007/BF02104515}{Commun.~Math.~Phys.~172,~187~(1995)}},
\texttt{\arxivref{hep-th/0303037}{hep-th/0303037}}.

\bibitem{Arzano:2021scz}
M.~Arzano and J.~Kowalski-Glikman,
\textit{``{Deformations of Spacetime Symmetries}: {Gravity, Group-Valued
  Momenta, and Non-Commutative Fields}''}.

\bibitem{Addazi:2021xuf}
A.~Addazi et~al.,
\textit{``{Quantum gravity phenomenology at the dawn of the multi-messenger
  era\textemdash{}A review}''},
\textsf{\doiref{10.1016/j.ppnp.2022.103948}{Prog.~Part.~Nucl.~Phys.~125,~103948~(2022)}},
\texttt{\arxivref{2111.05659}{arxiv:2111.05659}}.

\bibitem{Chu:1999gi}
C.-S.~Chu and P.-M.~Ho,
\textit{``{Constrained quantization of open string in background B field and
  noncommutative D-brane}''},
\textsf{\doiref{10.1016/S0550-3213(99)00685-9}{Nucl.~Phys.~B~568,~447~(2000)}},
\texttt{\arxivref{hep-th/9906192}{hep-th/9906192}}.

\bibitem{Schomerus:1999ug}
V.~Schomerus,
\textit{``{D-branes and deformation quantization}''},
\textsf{\doiref{10.1088/1126-6708/1999/06/030}{JHEP~9906,~030~(1999)}},
\texttt{\arxivref{hep-th/9903205}{hep-th/9903205}}.

\bibitem{Maldacena:1997re}
J.~M.~Maldacena,
\textit{``{The Large N limit of superconformal field theories and
  supergravity}''},
\textsf{\doiref{10.4310/ATMP.1998.v2.n2.a1}{Adv.~Theor.~Math.~Phys.~2,~231~(1998)}},
\texttt{\arxivref{hep-th/9711200}{hep-th/9711200}}.

\bibitem{Maldacena:1999mh}
J.~M.~Maldacena and J.~G.~Russo,
\textit{``{Large N limit of noncommutative gauge theories}''},
\textsf{\doiref{10.1088/1126-6708/1999/09/025}{JHEP~9909,~025~(1999)}},
\texttt{\arxivref{hep-th/9908134}{hep-th/9908134}}.

\bibitem{Hashimoto:1999ut}
A.~Hashimoto and N.~Itzhaki,
\textit{``{Noncommutative Yang-Mills and the AdS / CFT correspondence}''},
\textsf{\doiref{10.1016/S0370-2693(99)01037-0}{Phys.~Lett.~B~465,~142~(1999)}},
\texttt{\arxivref{hep-th/9907166}{hep-th/9907166}}.

\bibitem{Szabo:2001kg}
R.~J.~Szabo,
\textit{``{Quantum field theory on noncommutative spaces}''},
\textsf{\doiref{10.1016/S0370-1573(03)00059-0}{Phys.~Rept.~378,~207~(2003)}},
\texttt{\arxivref{hep-th/0109162}{hep-th/0109162}}.

\bibitem{Madore:2000en}
J.~Madore, S.~Schraml, P.~Schupp and J.~Wess,
\textit{``{Gauge theory on noncommutative spaces}''},
\textsf{\doiref{10.1007/s100520050012}{Eur.~Phys.~J.~C~16,~161~(2000)}},
\texttt{\arxivref{hep-th/0001203}{hep-th/0001203}}.

\bibitem{Kontsevich:1997vb}
M.~Kontsevich,
\textit{``{Deformation quantization of Poisson manifolds. 1.}''},
\textsf{\doiref{10.1023/B:MATH.0000027508.00421.bf}{Lett.~Math.~Phys.~66,~157~(2003)}},
\texttt{\arxivref{q-alg/9709040}{q-alg/9709040}}.

\bibitem{drinfeld_YBESolutions_1983}
V.~Drinfel'd,
\textit{``On constant quasi-classical solutions of the Yang-Baxter quantum
  equation''},
\textsf{Sov.~Math.~Dokl.~28~(1983)~667~66,~V.~Drinfel'd}.

\bibitem{Wess:2003da}
J.~Wess,
\textit{``{Deformed coordinate spaces: Derivatives}''},
\texttt{\arxivref{hep-th/0408080}{hep-th/0408080}},
in: \textit{``{1st Balkan Workshop on Mathematical, Theoretical and
  Phenomenological Challenges Beyond the Standard Model}: {Perspectives of
  Balkans Collaboration}''},
122--128p.

\bibitem{Chaichian:2004yh}
M.~Chaichian, P.~Presnajder and A.~Tureanu,
\textit{``{New concept of relativistic invariance in NC space-time: Twisted
  Poincare symmetry and its implications}''},
\textsf{\doiref{10.1103/PhysRevLett.94.151602}{Phys.~Rev.~Lett.~94,~151602~(2005)}},
\texttt{\arxivref{hep-th/0409096}{hep-th/0409096}}.

\bibitem{Chaichian:2004za}
M.~Chaichian, P.~P.~Kulish, K.~Nishijima and A.~Tureanu,
\textit{``{On a Lorentz-invariant interpretation of noncommutative space-time
  and its implications on noncommutative QFT}''},
\textsf{\doiref{10.1016/j.physletb.2004.10.045}{Phys.~Lett.~B~604,~98~(2004)}},
\texttt{\arxivref{hep-th/0408069}{hep-th/0408069}}.

\bibitem{Aschieri:2009zz}
P.~Aschieri, M.~Dimitrijevic, P.~Kulish, F.~Lizzi and J.~Wess,
\textit{``{Noncommutative spacetimes: Symmetries in noncommutative geometry and
  field theory}''}.

\bibitem{Vicedo:2015pna}
B.~Vicedo,
\textit{``{Deformed integrable {\ensuremath{\sigma}}-models, classical
  R-matrices and classical exchange algebra on Drinfel{\textquoteright}d
  doubles}''},
\textsf{\doiref{10.1088/1751-8113/48/35/355203}{J.~Phys.~A~48,~355203~(2015)}},
\texttt{\arxivref{1504.06303}{arxiv:1504.06303}}.

\bibitem{vanTongeren:2015uha}
S.~J.~van~Tongeren,
\textit{``{Yang{\textendash}Baxter deformations, AdS/CFT, and
  twist-noncommutative gauge theory}''},
\textsf{\doiref{10.1016/j.nuclphysb.2016.01.012}{Nucl.~Phys.~B~904,~148~(2016)}},
\texttt{\arxivref{1506.01023}{arxiv:1506.01023}}.

\bibitem{Borsato:2021fuy}
R.~Borsato, S.~Driezen and J.~L.~Miramontes,
\textit{``{Homogeneous Yang-Baxter deformations as undeformed yet twisted
  models}''},
\textsf{\doiref{10.1007/JHEP04(2022)053}{JHEP~2204,~053~(2022)}},
\texttt{\arxivref{2112.12025}{arxiv:2112.12025}}.

\bibitem{vanTongeren:2016eeb}
S.~J.~van~Tongeren,
\textit{``{Almost abelian twists and AdS/CFT}''},
\textsf{\doiref{10.1016/j.physletb.2016.12.002}{Phys.~Lett.~B~765,~344~(2017)}},
\texttt{\arxivref{1610.05677}{arxiv:1610.05677}}.

\bibitem{Meier:2025tjq}
T.~Meier and S.~J.~van~Tongeren,
\textit{``{Integrable spin chains in twisted maximally supersymmetric
  Yang-Mills theory}''},
\texttt{\arxivref{2507.18626}{arxiv:2507.18626}}.

\bibitem{Beisert:2003yb}
N.~Beisert and M.~Staudacher,
\textit{``{The N=4 SYM integrable super spin chain}''},
\textsf{\doiref{10.1016/j.nuclphysb.2003.08.015}{Nucl.~Phys.~B~670,~439~(2003)}},
\texttt{\arxivref{hep-th/0307042}{hep-th/0307042}}.

\bibitem{Minahan:2002ve}
J.~A.~Minahan and K.~Zarembo,
\textit{``{The Bethe ansatz for N=4 superYang-Mills}''},
\textsf{\doiref{10.1088/1126-6708/2003/03/013}{JHEP~0303,~013~(2003)}},
\texttt{\arxivref{hep-th/0212208}{hep-th/0212208}}.

\bibitem{Seiberg:1999vs}
N.~Seiberg and E.~Witten,
\textit{``{String theory and noncommutative geometry}''},
\textsf{\doiref{10.1088/1126-6708/1999/09/032}{JHEP~9909,~032~(1999)}},
\texttt{\arxivref{hep-th/9908142}{hep-th/9908142}}.

\bibitem{Douglas:2001ba}
M.~R.~Douglas and N.~A.~Nekrasov,
\textit{``{Noncommutative field theory}''},
\textsf{\doiref{10.1103/RevModPhys.73.977}{Rev.~Mod.~Phys.~73,~977~(2001)}},
\texttt{\arxivref{hep-th/0106048}{hep-th/0106048}}.

\bibitem{Guica:2017mtd}
M.~Guica, F.~Levkovich-Maslyuk and K.~Zarembo,
\textit{``{Integrability in dipole-deformed $\boldsymbol{\mathcal{N}=4}$ super
  Yang{\textendash}Mills}''},
\textsf{\doiref{10.1088/1751-8121/aa8491}{J.~Phys.~A~50,~39~(2017)}},
\texttt{\arxivref{1706.07957}{arxiv:1706.07957}}.

\bibitem{Lunin:2005jy}
O.~Lunin and J.~M.~Maldacena,
\textit{``{Deforming field theories with U(1) x U(1) global symmetry and their
  gravity duals}''},
\textsf{\doiref{10.1088/1126-6708/2005/05/033}{JHEP~0505,~033~(2005)}},
\texttt{\arxivref{hep-th/0502086}{hep-th/0502086}}.

\bibitem{Leigh:1995ep}
R.~G.~Leigh and M.~J.~Strassler,
\textit{``{Exactly marginal operators and duality in four-dimensional N=1
  supersymmetric gauge theory}''},
\textsf{\doiref{10.1016/0550-3213(95)00261-P}{Nucl.~Phys.~B~447,~95~(1995)}},
\texttt{\arxivref{hep-th/9503121}{hep-th/9503121}}.

\bibitem{Borowiec:2008uj}
A.~Borowiec and A.~Pachol,
\textit{``{kappa-Minkowski spacetime as the result of Jordanian twist
  deformation}''},
\textsf{\doiref{10.1103/PhysRevD.79.045012}{Phys.~Rev.~D~79,~045012~(2009)}},
\texttt{\arxivref{0812.0576}{arxiv:0812.0576}}.

\bibitem{Meier:2023lku}
T.~Meier and S.~J.~van~Tongeren,
\textit{``{Gauge theory on twist-noncommutative spaces}''},
\textsf{\doiref{10.1007/JHEP12(2023)045}{JHEP~2312,~045~(2023)}},
\texttt{\arxivref{2305.15470}{arxiv:2305.15470}}.

\bibitem{Aschieri:2006ye}
P.~Aschieri, M.~Dimitrijevic, F.~Meyer, S.~Schraml and J.~Wess,
\textit{``{Twisted gauge theories}''},
\textsf{\doiref{10.1007/s11005-006-0108-0}{Lett.~Math.~Phys.~78,~61~(2006)}},
\texttt{\arxivref{hep-th/0603024}{hep-th/0603024}}.

\bibitem{Dimitrijevic:2014dxa}
M.~Dimitrijevic, L.~Jonke and A.~Pachol,
\textit{``{Gauge Theory on Twisted $\kappa$-Minkowski: Old Problems and
  Possible Solutions}''},
\textsf{\doiref{10.3842/SIGMA.2014.063}{SIGMA~10,~063~(2014)}},
\texttt{\arxivref{1403.1857}{arxiv:1403.1857}}.

\bibitem{Hersent:2022gry}
K.~Hersent, P.~Mathieu and J.-C.~Wallet,
\textit{``{Gauge theories on quantum spaces}''},
\textsf{\doiref{10.1016/j.physrep.2023.03.002}{Phys.~Rept.~1014,~1~(2023)}},
\texttt{\arxivref{2210.11890}{arxiv:2210.11890}}.

\bibitem{Mathieu:2020ccc}
P.~Mathieu and J.-C.~Wallet,
\textit{``{Gauge theories on \ensuremath{\kappa}-Minkowski spaces: twist and
  modular operators}''},
\textsf{\doiref{10.1007/JHEP05(2020)112}{JHEP~2005,~112~(2020)}},
\texttt{\arxivref{2002.02309}{arxiv:2002.02309}}.

\bibitem{Meier:2023kzt}
T.~Meier and S.~J.~van~Tongeren,
\textit{``{Quadratic Twist-Noncommutative Gauge Theory}''},
\textsf{\doiref{10.1103/PhysRevLett.131.121603}{Phys.~Rev.~Lett.~131,~121603~(2023)}},
\texttt{\arxivref{2301.08757}{arxiv:2301.08757}}.

\bibitem{next-paper}
R.~Borsato and T.~Meier,
to appear.

\bibitem{Aschieri:2009ky}
P.~Aschieri and L.~Castellani,
\textit{``{Noncommutative D=4 gravity coupled to fermions}''},
\textsf{\doiref{10.1088/1126-6708/2009/06/086}{JHEP~0906,~086~(2009)}},
\texttt{\arxivref{0902.3817}{arxiv:0902.3817}}.

\bibitem{Aschieri:2005zs}
P.~Aschieri, M.~Dimitrijevic, F.~Meyer and J.~Wess,
\textit{``{Noncommutative geometry and gravity}''},
\textsf{\doiref{10.1088/0264-9381/23/6/005}{Class.~Quant.~Grav.~23,~1883~(2006)}},
\texttt{\arxivref{hep-th/0510059}{hep-th/0510059}}.

\bibitem{Meyer:1994wi}
U.~Meyer,
\textit{``{Wave equations on q Minkowski space}''},
\textsf{\doiref{10.1007/BF02101524}{Commun.~Math.~Phys.~174,~457~(1995)}},
\texttt{\arxivref{hep-th/9404054}{hep-th/9404054}}.

\bibitem{Majid:1994mh}
S.~Majid,
\textit{``{q epsilon tensor for quantum and braided spaces}''},
\textsf{\doiref{10.1063/1.531098}{J.~Math.~Phys.~36,~1991~(1995)}},
\texttt{\arxivref{hep-th/9406157}{hep-th/9406157}}.

\bibitem{Jurco:2001rq}
B.~Jurco, L.~Moller, S.~Schraml, P.~Schupp and J.~Wess,
\textit{``{Construction of nonAbelian gauge theories on noncommutative
  spaces}''},
\textsf{\doiref{10.1007/s100520100731}{Eur.~Phys.~J.~C~21,~383~(2001)}},
\texttt{\arxivref{hep-th/0104153}{hep-th/0104153}}.

\bibitem{Ciric:2021rhi}
M.~D.~\'Ciri\'c, G.~Giotopoulos, V.~Radovanovi\'c and R.~J.~Szabo,
\textit{``{Braided $L_{\infty}$-Algebras, Braided Field Theory and
  Noncommutative Gravity}''},
\texttt{\arxivref{2103.08939}{arxiv:2103.08939}}.

\bibitem{Giotopoulos:2021ieg}
G.~Giotopoulos and R.~J.~Szabo,
\textit{``{Braided symmetries in noncommutative field theory}''},
\textsf{\doiref{10.1088/1751-8121/ac5dad}{J.~Phys.~A~55,~353001~(2022)}},
\texttt{\arxivref{2112.00541}{arxiv:2112.00541}}.

\bibitem{Dimitrijevic:2011jg}
M.~Dimitrijevic and L.~Jonke,
\textit{``{A Twisted look on kappa-Minkowski: U(1) gauge theory}''},
\textsf{\doiref{10.1007/JHEP12(2011)080}{JHEP~1112,~080~(2011)}},
\texttt{\arxivref{1107.3475}{arxiv:1107.3475}}.

\bibitem{SpinorLie}
Y.~Choquet-Bruhat and C.~Dewitt-Morette,
\textit{``Analysis, Manifolds and Physics''},
North-Holland (2000),
Amsterdam,
433-524p.

\bibitem{FILK199653}
T.~Filk,
\textit{``Divergencies in a field theory on quantum space''},
\textsf{\doiref{https://doi.org/10.1016/0370-2693(96)00024-X}{Physics~Letters~B~376,~53~(1996)}},
\href{https://www.sciencedirect.com/science/article/pii/037026939600024X}{\texttt{https://www.sciencedirect.com/science/article/pii/037026939600024X}}.

\bibitem{Grosse:2005iz}
H.~Grosse and M.~Wohlgenannt,
\textit{``{On kappa-deformation and UV/IR mixing}''},
\textsf{\doiref{10.1016/j.nuclphysb.2006.05.004}{Nucl.~Phys.~B~748,~473~(2006)}},
\texttt{\arxivref{hep-th/0507030}{hep-th/0507030}}.

\bibitem{DimitrijevicCiric:2018blz}
M.~Dimitrijevic~Ciric, N.~Konjik, M.~A.~Kurkov, F.~Lizzi and P.~Vitale,
\textit{``{Noncommutative field theory from angular twist}''},
\textsf{\doiref{10.1103/PhysRevD.98.085011}{Phys.~Rev.~D~98,~085011~(2018)}},
\texttt{\arxivref{1806.06678}{arxiv:1806.06678}}.

\bibitem{Fokken:2013mza}
J.~Fokken, C.~Sieg and M.~Wilhelm,
\textit{``{The complete one-loop dilatation operator of planar real
  $\beta$-deformed $ \mathcal{N} $ = 4 SYM theory}''},
\textsf{\doiref{10.1007/JHEP07(2014)150}{JHEP~1407,~150~(2014)}},
\texttt{\arxivref{1312.2959}{arxiv:1312.2959}}.

\bibitem{Beisert:2005if}
N.~Beisert and R.~Roiban,
\textit{``{Beauty and the twist: The Bethe ansatz for twisted N=4 SYM}''},
\textsf{\doiref{10.1088/1126-6708/2005/08/039}{JHEP~0508,~039~(2005)}},
\texttt{\arxivref{hep-th/0505187}{hep-th/0505187}}.

\bibitem{vanTongeren:2019dlq}
S.~J.~van~Tongeren,
\textit{``{Unimodular jordanian deformations of integrable superstrings}''},
\textsf{\doiref{10.21468/SciPostPhys.7.1.011}{SciPost~Phys.~7,~011~(2019)}},
\texttt{\arxivref{1904.08892}{arxiv:1904.08892}}.

\bibitem{Borsato:2022ubq}
R.~Borsato and S.~Driezen,
\textit{``{All Jordanian deformations of the $AdS_5 \times S^5$
  superstring}''},
\textsf{\doiref{10.21468/SciPostPhys.14.6.160}{SciPost~Phys.~14,~160~(2023)}},
\texttt{\arxivref{2212.11269}{arxiv:2212.11269}}.

\bibitem{Borsato:2022drc}
R.~Borsato, S.~Driezen, J.~M.~Nieto~Garc{\'\i}a and L.~Wyss,
\textit{``{Semiclassical spectrum of a Jordanian deformation of
  AdS5{\texttimes}S5}''},
\textsf{\doiref{10.1103/PhysRevD.106.066015}{Phys.~Rev.~D~106,~066015~(2022)}},
\texttt{\arxivref{2207.14748}{arxiv:2207.14748}}.

\bibitem{Borsato:2024sru}
R.~Borsato and S.~Driezen,
\textit{``{Particle production in a light-cone gauge fixed Jordanian
  deformation of AdS5{\texttimes}S5}''},
\textsf{\doiref{10.1103/PhysRevD.111.086010}{Phys.~Rev.~D~111,~086010~(2025)}},
\texttt{\arxivref{2412.08411}{arxiv:2412.08411}}.

\bibitem{Borsato:2025smn}
R.~Borsato and M.~G.~Fern{\'a}ndez,
\textit{``{Jordanian deformation of the non-compact and $
  \mathfrak{s}{\mathfrak{l}}_2 $-invariant XXX$_{-1/2}$ spin-chain}''},
\textsf{\doiref{10.1007/JHEP08(2025)074}{JHEP~2508,~074~(2025)}},
\texttt{\arxivref{2503.24223}{arxiv:2503.24223}}.

\bibitem{deLeeuw:2025sfs}
M.~de~Leeuw, A.~Fontanella and J.~M.~Nieto~Garc{\'\i}a,
\textit{``{An integrable deformed Landau-Lifshitz model with particle
  production?}''},
\texttt{\arxivref{2506.13598}{arxiv:2506.13598}}.

\bibitem{Driezen:2025dww}
S.~Driezen and A.~Molines,
\textit{``{Jordanian spin chains for twisted strings in AdS5{\texttimes}S5}''},
\textsf{\doiref{10.1103/b1cg-6s5n}{Phys.~Rev.~D~112,~106001~(2025)}},
\texttt{\arxivref{2507.13911}{arxiv:2507.13911}}.

\bibitem{Driezen:2025izd}
S.~Driezen, F.~Levkovich-Maslyuk and A.~Molines,
\textit{``{Integrability for the spectrum of Jordanian AdS/CFT}''},
\texttt{\arxivref{2511.11521}{arxiv:2511.11521}}.

\bibitem{Beisert:2017pnr}
N.~Beisert, A.~Garus and M.~Rosso,
\textit{``{Yangian Symmetry and Integrability of Planar N=4 Supersymmetric
  Yang-Mills Theory}''},
\textsf{\doiref{10.1103/PhysRevLett.118.141603}{Phys.~Rev.~Lett.~118,~141603~(2017)}},
\texttt{\arxivref{1701.09162}{arxiv:1701.09162}}.

\bibitem{Beisert:2018zxs}
N.~Beisert, A.~Garus and M.~Rosso,
\textit{``{Yangian Symmetry for the Action of Planar $\mathcal N=$ 4 Super
  Yang-Mills and $\mathcal N=$ 6 Super Chern-Simons Theories}''},
\textsf{\doiref{10.1103/PhysRevD.98.046006}{Phys.~Rev.~D~98,~046006~(2018)}},
\texttt{\arxivref{1803.06310}{arxiv:1803.06310}}.

\bibitem{Garus:2018ajc}
A.~Garus,
\textit{``{Yangian Symmetry for the Action of the $N = 4$ Supersymmetric
  Yang{\textendash}Mills Theory}''}.

\bibitem{Beisert:2024wqq}
N.~Beisert and B.~K{\"o}nig,
\textit{``{Yangian form-alism for planar gauge theories}''},
\textsf{\doiref{10.1063/5.0253127}{J.~Math.~Phys.~66,~062301~(2025)}},
\texttt{\arxivref{2411.16176}{arxiv:2411.16176}}.

\bibitem{Beisert:2025mtn}
N.~Beisert and B.~K{\"o}nig,
\textit{``{Yangian symmetry escapes from the Fishnet}''},
\texttt{\arxivref{2511.19788}{arxiv:2511.19788}}.

\bibitem{Gurdogan:2015csr}
{\"O}.~G{\"u}rdo{\u{g}}an and V.~Kazakov,
\textit{``{New Integrable 4D Quantum Field Theories from Strongly Deformed
  Planar $\mathcal N = $ 4 Supersymmetric Yang-Mills Theory}''},
\textsf{\doiref{10.1103/PhysRevLett.117.201602}{Phys.~Rev.~Lett.~117,~201602~(2016)}},
\texttt{\arxivref{1512.06704}{arxiv:1512.06704}},
[Addendum: Phys.Rev.Lett. 117, 259903 (2016)].

\bibitem{Aharony:2008ug}
O.~Aharony, O.~Bergman, D.~L.~Jafferis and J.~Maldacena,
\textit{``{N=6 superconformal Chern-Simons-matter theories, M2-branes and their
  gravity duals}''},
\textsf{\doiref{10.1088/1126-6708/2008/10/091}{JHEP~0810,~091~(2008)}},
\texttt{\arxivref{0806.1218}{arxiv:0806.1218}}.

\bibitem{Hosomichi:2008jb}
K.~Hosomichi, K.-M.~Lee, S.~Lee, S.~Lee and J.~Park,
\textit{``{N=5,6 Superconformal Chern-Simons Theories and M2-branes on
  Orbifolds}''},
\textsf{\doiref{10.1088/1126-6708/2008/09/002}{JHEP~0809,~002~(2008)}},
\texttt{\arxivref{0806.4977}{arxiv:0806.4977}}.

\end{thebibliography}
\end{document}